\documentclass[10pt,journal]{IEEEtran}
\usepackage{booktabs}
\usepackage{multirow}
\usepackage{makecell}
\usepackage{amsthm}
\usepackage{diagbox}
\usepackage{xcolor}
\usepackage{subfigure}
\usepackage{amsmath,amssymb,amsfonts}
\usepackage{algorithm}
\usepackage{algorithmicx}
\usepackage{hyperref}
\usepackage{algpseudocode}
\usepackage{ragged2e}

\usepackage{graphicx}

\usepackage{epstopdf}

\newtheorem{myDef}{Definition}
\newtheorem{myTheo}{Theorem}
\newtheorem{myCoro}{Corollary}
\newtheorem{myRem}{Remark}

\ifCLASSOPTIONcompsoc
\usepackage[nocompress]{cite}
\else
  \usepackage{cite}
\fi
\ifCLASSINFOpdf
\else
   \usepackage[dvips]{graphicx}
\fi
\usepackage{array}
\begin{document}
%
\title{Auction-Promoted Trading for Multiple Federated Learning Services in UAV-Aided Networks}
%
%
%

\author{Zhipeng~Cheng,~\IEEEmembership{Student Member,~IEEE,} Minghui~Liwang,~\IEEEmembership{Member,~IEEE}, Xiaoyu~Xia,~\IEEEmembership{Member,~IEEE}, Minghui~Min,~\IEEEmembership{Member,~IEEE}, Xianbin Wang,~\IEEEmembership{Fellow,~IEEE,} Xiaojiang Du,~\IEEEmembership{Fellow,~IEEE}

\thanks{Zhipeng Cheng (chengzp\_x@163.com) and Minghui Liwang (Corresponding author, minghuilw@xmu.edu.cn) are with Department of Information and Communication Engineering, Xiamen University, Xiamen, China.  }

\thanks{Xiaoyu Xia (xiaoyu.xia@adelaide.edu.au) is with School of Computer Science, The University of Adelaide, Adelaide, Australia.}
\thanks{Minghui Min (minmh@cumt.edu.cn) is with School of Information and Control Engineering, China University of Mining and Technology, Xuzhou, China, and also with the Xuzhou First People's Hospital, Xuzhou, China.}
\thanks{Xianbin Wang (xianbin.wang@uwo.ca) is with Department of Electrical and Computer Engineering, Western University, Ontario, Canada.}
\thanks{Xiaojiang Du (dxj@ieee.org) is with Department of Electrical and Computer Engineering, Stevens Institute of Technology, Hoboken, NJ, USA.}
}

\markboth{IEEE Transactions on Vehicular Technology,~Vol.~XX, No.~XX, XXX~2015}
{}
\maketitle

\begin{abstract}
Federated learning (FL) represents a promising distributed machine learning paradigm that allows smart devices to collaboratively train a shared model via providing local data sets. However,  problems considering multiple co-existing FL services and different types of service providers are rarely studied. In this paper, we investigate a multiple FL service trading problem in Unmanned Aerial Vehicle (UAV)-aided networks, where FL service demanders (FLSDs) aim to purchase various data sets from feasible clients (smart devices, e.g., smartphones, smart vehicles), and model aggregation services from UAVs, to fulfill their requirements. An auction-based trading market is established to facilitate the trading among three parties, i.e., FLSDs acting as buyers, distributed located client groups acting as data-sellers, and UAVs acting as UAV-sellers. The proposed auction is formalized as a 0-1 integer programming problem, aiming to maximize the overall buyers' revenue via investigating winner determination and payment rule design. Specifically, since two seller types (data-sellers and UAV-sellers) are considered, an interesting idea integrating \textit{seller pair} and \textit{joint bid} is introduced, which turns diverse sellers into virtual seller pairs. Vickrey-Clarke-Groves (VCG)-based, and one-sided matching-based mechanisms are proposed, respectively, where the former achieves the optimal solutions, which, however, is computationally intractable. While the latter can obtain suboptimal solutions that approach to the optimal ones, with low computational complexity, especially upon considering a large number of participants. Significant properties such as truthfulness and individual rationality are comprehensively analyzed for both mechanisms. Extensive experimental results verify the properties and demonstrate that our proposed mechanisms outperform representative methods significantly.
\end{abstract}


\IEEEpeerreviewmaketitle

\section{Introduction}
\IEEEPARstart{T}{he} past decade has witnessed a rapid development of connected and intelligent society, mainly characterized by the deployment of advanced communication technologies (e.g., 4G/5G/B5G) and the proliferation of smart devices (e.g., smartphones, vehicles, edge servers) with enhanced computing/communication capabilities\cite{6GNetwk-1,new-add-zhang1}. Besides, the explosively growing big data generated by massive mobile devices have enabled a wide variety of innovative applications associated with artificial intelligence (AI) and machine learning (ML), e.g., E-health, smart city, intelligent transportation\cite{6GNetwk-2}. However, conventional AI and ML applications are generally conducted in a centralized manner, under which, data are frequently collected and transmitted from various mobile devices to centralized data centers, leading to overloaded wireless networks \cite{AliMag}. Meanwhile, the ever-growing concerns of data privacy and security also call for feasible and flexible distributed ML mechanisms. Regarding this backdrop, the interesting concept of federated learning (FL), firstly provided by Google, offers a privacy-preserving distributed ML paradigm and has attracted extensive attention from both industry and academia alike \cite{GoogleFL2016},\cite{FedCS-1}.

FL refers to the embodiment of ``bringing codes to data'', rather than ``bringing data to codes'' associated with traditional centralized ML \cite{GsongNet21}. For a typical FL scheme, a set of feasible mobile devices termed as clients or workers, will participate in the iterative training of FL. In each iteration round, every client first downloads a global model from a centralized aggregator, and then, conducts local training based on its local data set, i.e, ML-as-a-service (MLaaS) \cite{LaaS}. Finally, the model parameters (or gradients) are uploaded to the aggregator synchronously (or asynchronously) and aggregated as a new global model via a specific aggregation algorithm, e.g., FedAvg \cite{FedAvg}. Recently, extensive efforts are devoted to different aspects of FL, e.g., client selection \cite{FedCS-1, FedCS-2}, resource management \cite{FedRM-1, FedRM-2}, architecture design \cite{FedHier-1,FedHier-2} and incentive mechanism design \cite{GsongNet21}, \cite{FedIncen-1}. Although existing works have made lots of contributions, most of them focus on the single FL service scenario, while neglecting multiple co-existing FL services. Since many smart devices are embedded with various sensors that can collect different data types for diverse FL services, which thus support multiple co-existing FL services to be trained over the same group of clients, and presents a novel problem that is worth attentions\cite{MultiFL-IOT, MultiFL-TWC, MultiFL-TMC}.

Generally, different FL services could have different service requirements and economic values, where local data qualities (e.g., data size, data distribution) of clients are significant in evaluating FL training performance\cite{FedIncen-1}. Besides, many commercial or academic FL services often suffer from the lack of training data and participating clients, which thus calls for an efficient trading ecosystem, to facilitate the practical implementation of FL services\cite{DataTrade-TIFS}. Thus, selecting feasible clients for different FL services according to clients' data qualities is essential to guarantee mutual success, while maximizing the overall economic values. Nevertheless, clients are usually reluctant to provide data without profits due to selfishness, which necessitates proper incentive mechanism design. To fully motivate clients to participate in FL training and report true data profiles, it is critical to provide appropriate monetary reward for its involvement. Recently, auction theory has been adopted as a popular incentive mechanism thanks to its fairness and trading efficiency, where participants can buy/sell requested services/resources at reasonable prices\cite{NewaddTPDS-1,NewaddTPDS-2,NewaddTPDS-3,New-review3}. To this end, auction becomes an effective method to facilitate a trading market between FL services and clients, where FL service demanders (FLSDs) act as buyers; while clients act as sellers to provide data sets and MLaaS \cite{MultiFL-TWC},\cite{Auction-FL-TWC}.

One common challenge faced by FL services in wireless communication networks is the inherent communication cost (e.g., delay, energy consumption, bandwidth), mainly incurred during FL model uploading/downloading, especially when facing unstable communication links. Thus, it is necessary to utilize an edge server as an intermediate FL model aggregator to facilitate communication-efficient FL services\cite{FedHier-1}. Particularly, Unmanned Aerial Vehicle (UAV) is introduced as a convenient platform for FL due to its flexibility\cite{new-add-zhang2,FL-UAV-Network,21-TITS-MC,TITS-A-C}, performing as a model aggregator and relay. Similar to the clients, monetary rewards should be paid to UAVs to cover the overhead (e.g., delay and energy consumption) incurred during service provisioning, as the major motivation of its participation in the FL service market. Apparently, UAVs can also be regarded as sellers that can offer model aggregation and delivery services. As a result, in a UAV-aided multiple FL service trading market, determining applicable clients and UAVs for different FL services under appropriate incentives becomes urgent and critical.

In this paper, we investigate an auction-promoted multiple FL service trading problem in UAV-aided networks, where multiple FLSDs act as buyers to purchase data sets from groups of clients (data-sellers), as well as model aggregation and delivery services from UAVs (UAV-sellers), to fulfill their FL service requirements. To the best of our knowledge, this paper is among the first to establish an efficient FL service trading market among three parties: FLSDs, data-sellers, and UAV-sellers, relying on a well-designed reverse auction. More importantly, significant auction properties such as truthfulness, individual rationality, and computational efficiency are also carefully analyzed. Main contributions of this paper are summarized as follows:
\begin{itemize}
  \item We propose a reverse auction model for multiple FL service trading market under UAV-aided networks via considering three different parties: multiple FLSDs as buyers; client groups as data sellers, and UAVs as UAV-sellers. Specifically, the buyers and sellers submit their bid profiles to an auctioneer who manages the auction process. The proposed auction is formalized as a 0-1 integer programming problem aiming to maximize the overall buyers' revenue, where the winner determination problem and payment rule among FLSDs, data sellers, and UAV sellers are carefully analyzed.

  \item We first study a Vickrey-Clarke-Groves (VCG)-based reverse auction mechanism that can obtain the optimal solutions, which guarantees the truthfulness and individual rationality of sellers. Specifically, since two seller types (data-sellers and UAV-sellers) are considered in our proposed auction model, an interesting idea of \textit{seller pair} and \textit{joint bid} is introduced to facilitate the auction procedure by turning each seller into multiple virtual sellers.

  \item The exponential computational complexity for obtaining optimal auction solutions by VCG-based mechanism poses great challenges (particularly upon considering a large number of participants). We thus propose a computation-efficient suboptimal auction mechanism, based on one-sided matching, while properties associated with which have also been comprehensively proved.

  \item We conduct extensive simulations to evaluate the performance and verify the properties of our proposed mechanisms. Experimental results demonstrate the commendable performance gain of our proposed mechanisms, in comparison with baseline methods.
\end{itemize}
To achieve a better understanding of both the problem and roles of participants in the proposed auction, online shopping in real-life is illustrated as an interesting example. In online shopping, different buyers have different demands on various commodities, such as purchasing a refrigerator or air conditioner from an online store. In this case, buyers need to select a feasible store according to their own needs (e.g., budget,) and stores' bidding prices. Generally, the required commodity needs to be delivered by an express company from the chosen store, where different express companies may offer different service qualities (e.g., estimated delivery time) by charging different fees according to commodity characteristics, e.g., delivery distance, commodity weight. Consequently, each buyer has to determine the following key problem: which commodity in which store via which express company, upon considering lots of factors, to minimize their costs. Interestingly, our proposed auction model can refer to the above-mentioned example, where online stores and express companies are regarded as data-sellers and UAV-sellers, respectively, while data sets and UAV services are seen as commodities.

The rest of this paper is structured as follows. Related works are discussed in Section \uppercase\expandafter{\romannumeral2}. System model and problem formulation are presented in Section \uppercase\expandafter{\romannumeral3}. VCG-based optimal reverse auction and the computation-efficient suboptimal reverse auction is proposed in Section \uppercase\expandafter{\romannumeral4} and \uppercase\expandafter{\romannumeral5}, respectively. Numerical simulations are conducted in Section \uppercase\expandafter{\romannumeral6} before we conclude the paper in Section \uppercase\expandafter{\romannumeral7}.
%

%

\section{Related Work}
Although many existing works have been devoted to investigating different challenges under considering single FL service, e.g., some of them have been classified and summarized in \cite{FLSurvey-1, FLSurvey-2, FLSurvey-3}, only a few pioneering works put efforts to multiple FL services, such as \cite{MultiFL-IOT, MultiFL-TWC, MultiFL-TMC}. In \cite{MultiFL-IOT}, each FL service is regarded as an FL task and the authors propose a many-to-many two-sided matching algorithm for Multitask FL. However, they only consider how to match the clients to FL tasks at preset edge servers to minimize the overall training latency of one global aggregation round. \cite{MultiFL-TWC} studies the bandwidth allocation problem for multiple simultaneous FL services where the clients of each FL service share the common bandwidth resources. The authors propose a two-level resource allocation framework to tackle the intra- and inter-service bandwidth allocation problem. \cite{MultiFL-TMC} investigates a general resource sharing problem for multiple FL services, different from \cite{MultiFL-IOT}, \cite{MultiFL-TWC} and our work, the authors assume each client can participate in the training of multiple FL services simultaneously, and thus not only the communication resources are shared by clients of each FL service but also the local computation (i.e., CPU) resources of each client are allocated among multiple FL services. However, none of these previous works consider the data and service trading problem among FLSDs, client groups, and UAVs.

Auction theory has been proven as an efficient incentive mechanism for FL regarding economic optimality and properties\cite{Auction-FL-TWC},\cite{TITS-A-C},\cite{TMCAuction}. In \cite{Auction-FL-TWC}, an auction approach is proposed as an incentive mechanism for FL in a wireless cellular network that involves one base station and multiple clients, where the base station acts as an auctioneer and the clients are the sellers. A primal-dual auction mechanism is proposed to decide the winners and maximize social welfare. The authors in \cite{TITS-A-C} study a communication-efficient FL in a UAV-enabled vehicular network, where the UAVs act as a wireless relay to facilitate the communication between clients and FL aggregator. A joint auction-coalition framework is proposed to allocate UAVs to different coalitions to maximize UAVs' profits. \cite{TMCAuction} proposes an auction framework to incentivize clients to provide local data sets for FL service. The authors first propose an approximate strategy-proof auction mechanism to maximize social welfare, then an automated auction framework is further proposed to improve social welfare based on graph neural networks and deep reinforcement learning. Although this work also considers the data trading between the clients and FL services, only a single FL service is considered.

UAVs have been widely applied in FL serving as FL aggregators, wireless relays, or clients. \cite{FL-UAV-Network} discusses the UAV-assisted FL where the FLSD can employ UAVs as intermediate model aggregation servers and mobile relays to deliver models so as to increase the reach of FL and improve communication efficiency. A multi-dimensional contract mechanism is proposed to incentivize the UAVs as a case study. Similarly, \cite{21-TITS-MC} also proposes a multi-dimensional contract matching approach in a UAV-enabled vehicular network where UAVs are clients to conduct data collection and model training for FLSD. \cite{FL-UAV-TVT} considers FL-aided image classification tasks in UAV-aided exploration scenarios where the UAVs act as clients to collect image data and train a classification model. However, none of these works have considered UAVs as sellers for multiple FL services trading market, which represents the major difference and novelty in our paper.

\section{System Model and Problem Formulation}
This section introduces the auction model for the multiple FL service trading market, the auction properties, and the problem formulation.
\begin{figure}[t]
	\centering
	\includegraphics[width=3.4in]{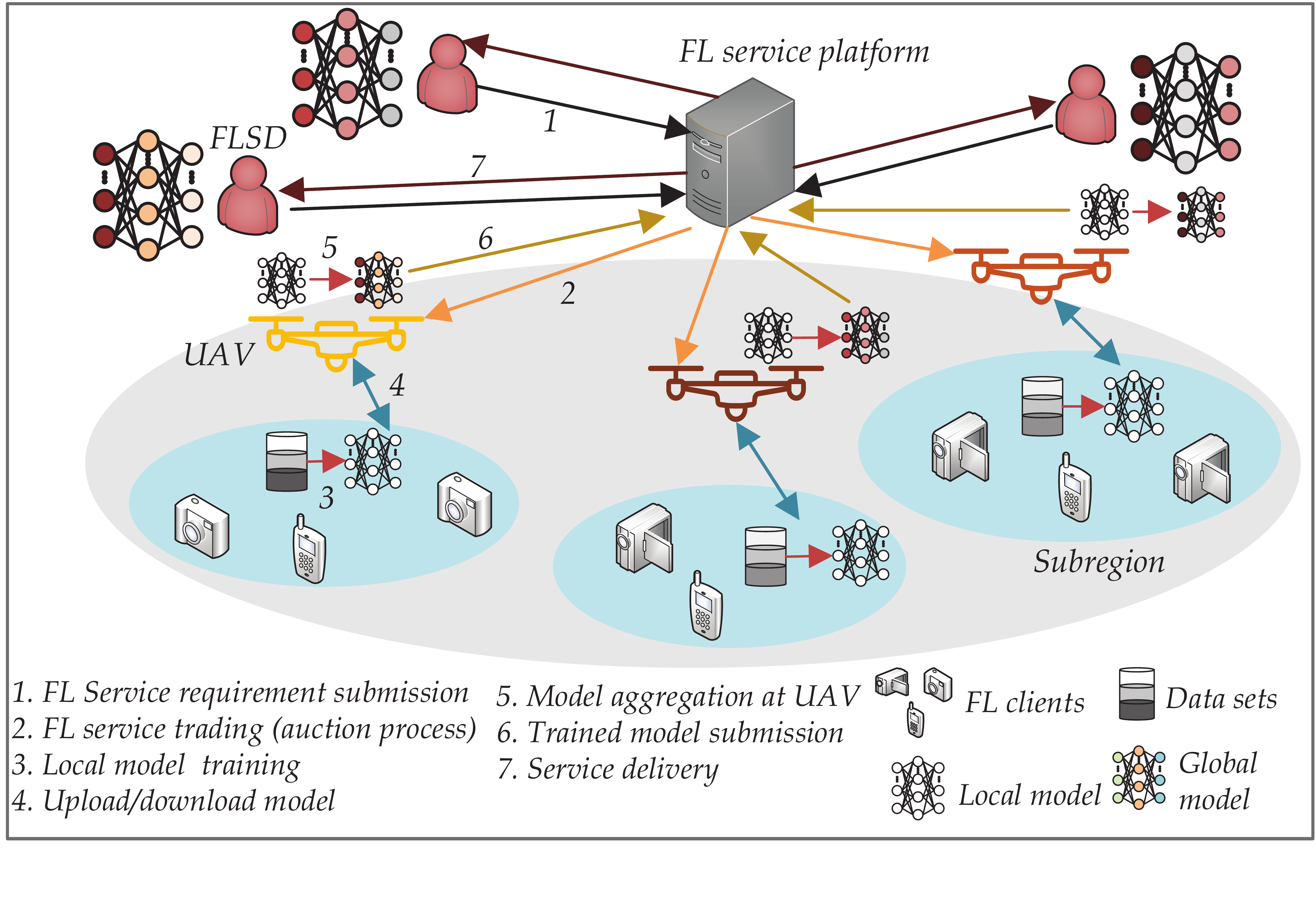}
	\caption{An illustration of Multiple FL service trading process in UAV-aided networks.}
	\label{fig1}
\end{figure}

\subsection{Overview of the proposed auction}
As depicted in Fig. \ref{fig1}, we consider a scenario of multiple co-existing FL services in a UAV-aided network, which contains multiple co-existing FLSDs, geographically isolated subregions (each of which is composed of a set of clients such as smart phones, smart vehicles) and UAVs. Specifically, each FLSD has an FL service that requires to train a unique ML model, which, however, suffers from the lack of training data mainly owing to privacy concerns. Thus, FLSDs have to train their models under an FL manner by purchasing training data and MLaaS from subregions. Additionally, assuming that each subregion consists of a sufficient number of homogeneous smart devices to cooperatively participate in the market, for collecting data and training ML models, under a consensus mechanism\cite{FedIncen-1}. To this end, each subregion can be regarded as a hyper data owner (DO), where terms ``subregion'' and ``DO'' are interchangeable with each other hereafter, for analytical simplicity.

Due to the selfishness of participants, proper incentives are encouraged. To compensate for the cost of clients (e.g., data collection cost, storage cost, model training cost), DOs can sell their data sets to FLSDs by charging corresponding fees. To bridge the FLSDs' service requirements and DOs' data sets in such an open market, auction has been emerged as a feasible paradigm to incentivize both sides. In our proposed auction model, FLSDs can report their requirements, while DOs can submit the description of their data sets and sell-bids, where corresponding FL services can be assigned to feasible DOs.

However, FL training may incur substantial cost on both computation and communication, especially for FL model transmission cost, which includes model downloading/uploading cost between FLSDs and DOs in dynamic wireless networks, etc. For example, unstable communication links among FLSDs and DOs greatly call for flexible intermediates to facilitate communication-efficient FL training, especially for large ML models. Thus, in the proposed auction, UAVs are serving as model aggregators for local trained model aggregation and model delivery, as shown in Fig. \ref{fig1}. When an FL service is assigned to a DO, a feasible UAV can be employed which flies to the DO from the UAV base, and hovers over the corresponding region to serve clients until the FL training is finished. Consequently, FLSDs also have to pay for the time and energy cost for services offered by UAVs. Since different UAVs have different capabilities (e.g., communication/computation ability, flying speed) and costs, each FLSD has to be mapped to a suitable UAV to promote the completion of FL training and minimize the overall payments. For analytical simplicity, assume that each FLSD can only buy the data set from one DO and employ one UAV in this paper.\footnote{Note that our proposed auction can also be well applied in scenarios where each FLSD can select multiple DOs or UAVs. Under this circumstance, each DO (or UAV) can be seen as multiple DOs, regarding different FLSDs' demands. This topic will be studied as an interesting future work.}

\begin{figure}[t]
	\centering
	\includegraphics[width=3.4in, height=1.5in]{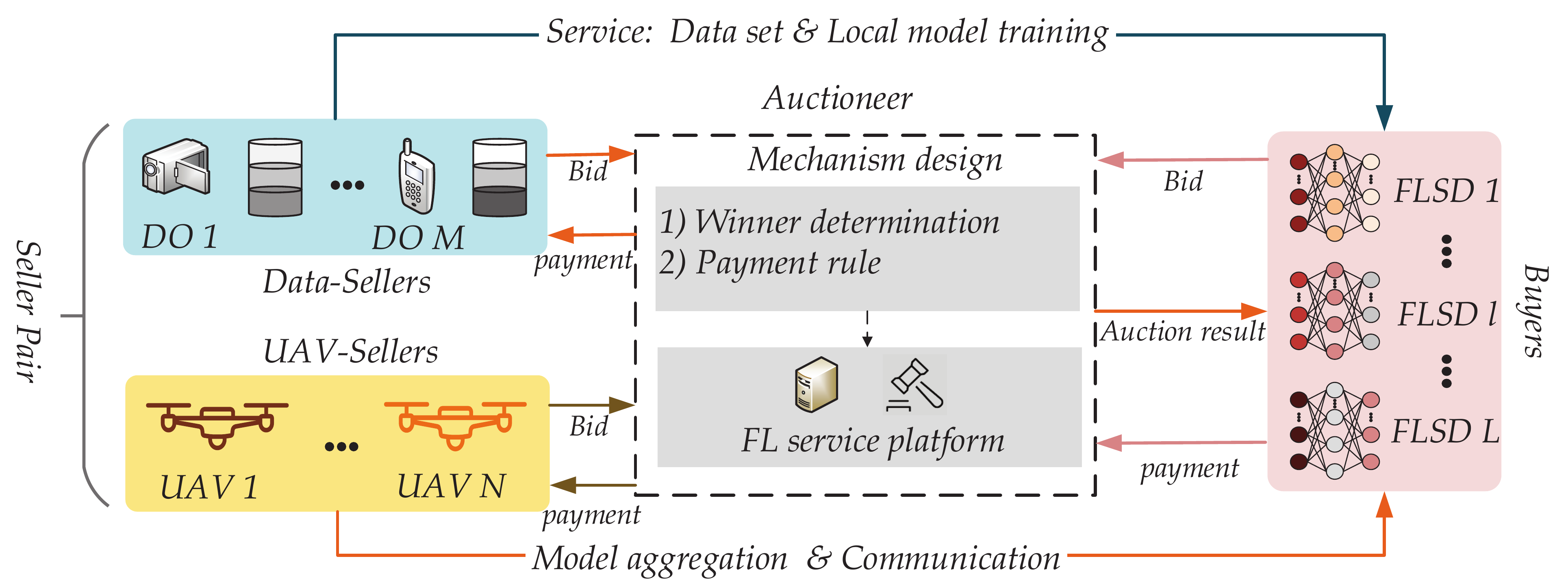}
	\caption{ Multiple FL services trading based on the proposed reverse auction model.}
	\label{fig2}
\end{figure}

To guarantee the auction efficiency and trading stability among the FLSDs, DOs and UAVs, a trustworthy third party (e.g., base station, road side unit, etc) is introduced as a auctioneer to manage the auction process. Fig. \ref{fig2} illustrates the proposed auction-based FL service market. Major notations and definitions are summarized in Table \ref{tab1}.
\subsection{Auction participants}
\subsubsection{Data-sellers}
As mentioned previously, the considered region is divided into a set of $M$ geographically isolated DOs, denoted by $\mathcal{M}= \{1,\dots,m,\dots, M\}$, e.g., through a clustering or partitioning algorithm \cite{21-TITS-MC}. For analytical simplicity, we assume that each DO $m \in \mathcal{M}$ has a set of identically distributed clients, where each of which is equipped with different sensors (e.g., camera, temperature sensor, acceleration sensor), and thus can collect different training data sets (types) for different ML models. Suppose that clients can collect $L$ types of data sets associated with $L$ ML models. Generally, training performance of an ML model can be approximated by the collective data quality (e.g., overall \textit{data size} and \textit{data distribution} \cite{FedIncen-1},\cite{TMCAuction}). For example, the relationship between the test accuracy $\Delta$ and data size $d$ on the MNIST data set can be fitted into a concave function\cite{FedIncen-1}, given by the following (\ref{e1}),
\begin{equation}\label{e1}
\Delta=\alpha_{1}*log(1+\alpha_{2}*d),
\end{equation}
where $\alpha_{1}$ and $\alpha_{2}$ are positive curve fitting parameters.

In this paper, we mainly consider the overall data size of a DO as the indicator of collective data quality (complicated data quality parameters will be considered in our future work), where the data size vector of DO $m$ regarding $L$ data sets is denoted by $\vec{d}_{m}= (d_{m,1},\dots,d_{m,l},\dots,d_{m,L})$. Notably, to ease concerns about data distribution, we assume the data sets of DOs are independent and identically distributed, and can cover all the data classes for each FLSD. When DO $m$ participates in the auction, it submits a bundle $B_{m}^{d}=<\vec{d}_{m},\vec{q}_{m}>$ to the auctioneer, where $\vec{q}_{m}=(q_{m,1},\dots,q_{m,l},\dots,q_{m,L})$ represents the sell-bid vector of DO $m$ and $L$ data sets. Moreover, let $\vec{c}_{m}=(c_{m,1},\dots,c_{m,l},\dots,c_{m,L})$ denote the corresponding data cost vector, where ${c}_{m,l}=\sigma_{m,l}d_{m,l}$ describes the data cost of data set $l$, and $\sigma_{m,l}$ represents the average unit data cost (and the local training cost) of all the clients in DO $m$, which mainly indicates the differences between different data set types and data-seller types. The data costs mainly rely on each DO's private values and are unavailable to others.

\begin{table}[t]\label{tab1}
\newcommand{\tabincell}[2]{\begin{tabular}{@{}#1@{}}#2\end{tabular}}
\centering
\caption{Major Notations and Definitions}
\label{tab1}
\setlength{\tabcolsep}{0.0mm}{
\begin{tabular}{|c|l|}
\hline
Notation&\tabincell{l}{Definition}\\
\hline
$M/N/L$&\tabincell{l}{Number of data-sellers/UAV-sellers/Buyers}\\
\hline
$\mathcal{M}/\mathcal{N}/\mathcal{L}$&\tabincell{l}{Set of data-sellers/UAV-sellers/Buyers}\\
\hline
$\vec{d}_{m}$&\tabincell{l}{Data size vector of data-seller $m$}\\
\hline
$q_{m,l}$&\tabincell{l}{Sell-bid of data-seller $m$ for buyer $l$}\\
\hline
$c_{m,l}$&\tabincell{l}{Data cost of data-seller $m$ for buyer $l$}\\
\hline
$\sigma_{m,l}$&\tabincell{l}{Unit data cost of data-seller $m$ for buyer $l$}\\
\hline
$e^{f}_{n,m}$&\tabincell{l}{Flying cost of UAV-seller $n$ to data-seller $m$}\\
\hline
$\lambda_{n}$&\tabincell{l}{Unit flying cost of UAV-seller $n$}\\
\hline
$t_{n,m}$&\tabincell{l}{Distance between UAV-seller $n$ and data-seller $m$}\\
\hline
$e^{s}_{n,l}$&\tabincell{l}{Service cost of UAV-seller $n$ for buyer $l$}\\
\hline
$e_{n,(m,l)}$&\tabincell{l}{Total cost of UAV-seller $n$}\\
\hline
$s_{n,m,l}$&\tabincell{l}{Sell bid of UAV-seller $n$}\\
\hline
$v_{l,(m,n)}$&\tabincell{l}{Valuation of buyer $l$ with seller pair $(m,n)$ }\\
\hline
$x_{l,m,n}$&\tabincell{l}{Winner determination decision 0-1 variable}\\
\hline
$\mathbf{X}$&\tabincell{l}{Winner determination decision matrix}\\
\hline
$p^{d}_{m}$&\tabincell{l}{Payment of data-seller $m$}\\
\hline
$p^{u}_{n}$&\tabincell{l}{Payment of UAV-seller $n$}\\
\hline
$U^{d}_{m}/U^{u}_{n}/U^{b}_{l}$&\tabincell{l}{Revenue of data-seller $m$/UAV-seller $n$/buyer $l$}\\
\hline
$\mathcal{M}_{l}$&\tabincell{l}{Virtual data-seller set for buyer $l$}\\
\hline
$\mathcal{N}_{l,m}$&\tabincell{l}{Virtual UAV-seller set for buyer $l$}\\
\hline
$J_{l,(m,n)}$&\tabincell{l}{Joint bid of seller pair $(m,n)$ for buyer $l$}\\
\hline
$P^{f}_{m,n}$&\tabincell{l}{Total payment of seller pair $(m,n)$}\\
\hline
$\mathbf{T_{A}}/\mathbf{T_{l}}$&\tabincell{l}{Preference list of auctioneer/ buyer $l$}\\
\hline
\end{tabular}}
\end{table}
\subsubsection{UAV-sellers}
Considering a set $\mathcal{N}=\{1,2,\dots,n,\dots,N\}$ of $N$ independent UAVs, where each UAV $n \in \mathcal{N}$ can be regarded as a UAV-seller\footnote{Note that each UAV $n$ can be regarded as a single UAV or a group of UAVs of the same type.}. Once a UAV is employed to serve a DO, costs are incurred, which comprises two key parts: UAV flying cost $\vec{e}^{f}_{n}$ and service cost $\vec{e}^{s}_{n}$. The flying cost accounts for the travel delay and energy cost when UAV $n$ flies to DO $m$ before the training starts, depending on the traversal distance $t_{n,m}$ between its starting point and DO $m$. Thus, the flying cost $e^{f}_{n,m}\in \vec{e}^{f}_{n}$ of UAV $n$ for DO $m$ is calculated by the following (\ref{e2}),
\begin{equation}\label{e2}
e^{f}_{n,m} =\lambda_{n}t_{n,m},
\end{equation}
where $\lambda_{n}$ is the unit flying cost (e.g., unit time cost, unit energy cost) associated with UAV $n$.

Service cost mainly concerns the cost during the UAV service period (e.g., communication
cost, UAV hovering cost). Since different ML models have different model sizes and training requirements, the service cost mainly relies on the type of learning model. For analytical simplicity, the trivial details of UAV communications, such as model uploading/ broadcasting are omitted in the proposed auction (more details can refer to existing works, e.g., \cite{TITS-A-C}). The service cost of UAV $n$ for FLSD $l$ can be quantified by the following (\ref{e3}),
\begin{equation}\label{e3}
e^{s}_{n,l}=\mathbf{f}(\Theta_{n,l}),
\end{equation}
where $\mathbf{f}(\cdot)$ represents a monotone function that determines the service cost of UAV in different scenarios\footnote{Note that service cost involves factors such as delay or energy consumption of UAV and clients during the UAV service period, while this paper puts less emphasis on the exact models of delay and energy consumption for the following two reasons. First, we pay less attention on any specific cost functions since  a general quantitative cost is sufficient for the auction process, as also supported by existing works\cite{FedIncen-1,TMC20-BudgetConstraint, LiwangIoT2019, U-cost-1,U-cost-2}. Second, our proposed mechanism can also be well applied when considering detailed delay and energy consumption models as proposed in \cite{MultiFL-IOT} and \cite{MultiFL-TWC}, although there may be some differences associated with mathematical derivations (similar ideas can be found in \cite{LiwangTWC22})}. $\Theta_{n,l}$ indicates the properties of UAV $n$ when serving FLSD $l$, such as the computing ability (e.g., CPU, GPU), the communication data rate and reliability. This function can be altered according to specific scenarios. Correspondingly, the total cost for UAV $n$ to support the training of FL service $l$ with DO $m$ can be denoted as,
\begin{equation}\label{e4}
e_{n,(m,l)}=e^{f}_{n,m}+e^{s}_{n,l},
\end{equation}
thus, the total cost of UAV-seller $n$ can be denoted as a matrix $e_{n,(m,l)} \in \mathbf{E}_{n}$, which captures the bidirectional dependence of UAV-sellers.

Consequently, when UAV $n$ participates in the market, it submits a bundle $B_{n}^{u}=<\vec{t},\vec{\Theta}_{n},\mathbf{S_{n}} >$ to the auctioneer, where $\vec{t}$ is the traversal distance vector of $M$ DOs, and $\vec{\Theta}_{n}$ represents the general service capability vector. And $\mathbf{S_{n}}$ is the sell-bid matrix of size $M*L$, where $s_{n,m,l} \in \mathbf{S}_{n}$  stands for the sell-bid of UAV $n$ to serve FLSD $l$ with DO $m$.

\subsubsection{Buyers}
Considering a set of $L$ disparate FLSDs in the proposed market, as denoted by $\mathcal{L}=\{1,\dots,l,\dots, L\}$. Each FLSD $l$\footnote{We use $l$ to denote a FLSD, the FL service of the corresponding FLSD, the corresponding data type of a DO and a buyer in this paper.} acts as a buyer that requires to train an FL model $l$ to fulfill its service requirement. Similarly, buyer $l$ submits the corresponding bid information $d_{l}$ to the auctioneer, where $d_{l}$ denotes the corresponding service requirement, which is the minimum data size that can satisfy the service requirement\footnote{We assume $d_{l}$ can be estimated according to a similar function as Eq. (1), however, it is out of the scope of this paper to give the detail functions for different ML models.}. Since different DOs and UAVs can bring different service qualities, each buyer $l$ holds a private value matrix $\mathbf{V}_{l}=\{v_{l,(m,n)}|m \in \mathcal{M}, n\ \in \mathcal{N}\}$, where $v_{l,(m,n)}$ indicates the private valuation which represents the monetary reward obtained by $l$ when the service requirement is met by sellers $m$ and $n$; while $v_{l,(m,n)}=0$, otherwise. Then, the valuation of buyer $l$ is defined as (\ref{e5}),
\begin{equation}\label{e5}
  v_{l,(m,n)}=\Psi(d_{m,l},\Theta_{n,l}),
\end{equation}
where $\Psi(\cdotp)$ represents a general monotonic function.
\subsubsection{Auctioneer}
As shown in Fig. \ref{fig2}, the proposed auction process can be managed by the auctioneer via solving the following key problems:
\begin{itemize}
  \item \textit{Winner determination (FLSD-DO-UAV matching)}: After receiving the sell-bids from data-sellers and UAV-sellers as well as service requirements from buyers, the auctioneer initializes the auction process and decides winning buyers and winning sellers. Specifically, the auctioneer matches each winning buyer to applicable winning data-seller and UAV-seller (one-to-one-to-one matching), i.e., $x_{l,m,n}:\{l: l\in \mathcal{L}\}{\rightarrow} \{m: m\in \mathcal{M}\}{\rightarrow} \{n: n\in \mathcal{N}\}$. Specifically, $x_{l,m,n}$ represents the binary winner determination variable, where $x_{l,m,n}=1$ when buyer $l$ wins the auction and can be matched to winning sellers $m$ and $n$; otherwise, $x_{l,m,n,}=0$.

  \item \textit{Payment rule}: The auctioneer further determines the final payments for winning sellers from corresponding buyers with a feasible payment rule, namely, the final payment profiles of the sellers. Let $\vec{P}^{d}=<p^{d}_{1},\dots,p^{d}_{m},\dots,p^{d}_{M}>$ and $\vec{P}^{u}=<p^{u}_{1},\dots,p^{u}_{n},\dots,p^{u}_{N}>$ denote the final payment profiles of $M$ data-sellers and $N$ UAV-sellers, respectively. Apparently, buyers/sellers pay/receive nothing if they fail in the proposed auction. Besides, to promote the payment transaction between buyers and sellers in practice, a centralized credit-based transaction management system (e.g., as presented in \cite{New-review3} and \cite{INFOCOM03}), can be adopted. Specifically, all the buyers and sellers can have bank accounts in a dedicated band entity, and when the trading among the buyers and the sellers is over, the buyers can transfer payments to the designated sellers' accounts through the bank. Moreover, the innovative development of electronic payment makes the transaction management more convenient.
\end{itemize}

\subsection{Revenues and economical properties}
Key revenue functions and properties associated with the proposed auction are introduced hereafter.
\subsubsection{Revenues of participants}
\begin{myDef}
(Revenue of data-seller): The revenue of data-seller $m$ is defined as the difference between the final received payment $p_{m}^{d}$ and the bid ${q}_{m,l}$, i.e.,
 \begin{equation}\label{e6}
U^{d}_{m}= \sum \limits_{l \in \mathcal{L}}\sum \limits_{n \in \mathcal{N}}x_{l,m,n}(p_{m}^{d}-{q}_{m,l}).
\end{equation}
\end{myDef}


\begin{myDef}
(Revenue of UAV-seller): The revenue of UAV-seller $n$ is defined by the difference between the final received payment $p_{n}^{u}$ and the bid ${s}_{n,(m,l)}$, i.e.,
\begin{equation}\label{e7}
U^{u}_{n}=  \sum \limits_{l \in \mathcal{L}}\sum \limits_{m \in \mathcal{M}}x_{l,m,n}(p_{n}^{u}-{s}_{n,(m,l)}).
\end{equation}
\end{myDef}


\begin{myDef}
(Revenue of buyer): The revenue of buyer $l$ is defined as the difference between the valuation ${v}_{l,(m,n)}$ and total bid $({q}_{m,l}+{s}_{n,(m,l)})$, which can be regarded as the maximum gain brought by the trading, i.e.,
\begin{equation}\label{e8}
U_{b}^{l}= \sum \limits_{m \in \mathcal{M}} \sum \limits_{n \in \mathcal{N}}x_{l,m,n}\left({v}_{l,(m,n)}-({q}_{m,l}+{s}_{n,(m,l)})\right).
\end{equation}
\end{myDef}





\subsubsection{Economical properties}
In this paper, we consider a single-round and sealed-bid auction without collusion among the auction participants. Major properties considered in the auction mechanism design are :
\begin{itemize}
  \item \textit{Truthfulness (Incentive Compatibility).} An auction is truthful (incentive compatible) if all the auction participants can maximize their revenues when bidding truthfully, i.e., for each data-seller $m$ and UAV-seller $n$, $q_{m,l}= c_{m,l}$ and $s_{n,(m,l)}= e_{n,(m,l)}$ holds, and none of them can increase their revenues by misreporting the bid information. For example, if data-seller $m$ gets revenue $U_{m}^{d}$ and $\overline{U}_{m}^{d}$ for bidding truthfully and untruthfully, respectively, $U^{d}_{m}\geq \overline{U}^{d}_{m}$ always holds in a truthful auction.
  \item \textit{Individual Rationality.} An auction is individual rational if all the winning sellers and buyers can get non-negative revenues. For example, if data-seller $m$ wins the auction, then $U^{d}_{m}\geq 0$ always holds in a individually rational auction. In addition, if an auction is both truthful and individual rational, then the auction is \textit{strategy-proof}\cite{STAR}.
  \item \textit{Computational Efficiency.} An auction is computationally efficient if the auction results (i.e., winner determination and final payment profiles) can be obtained within polynomial time.
\end{itemize}

\subsection{Problem formulation}
In this paper, the proposed auction mechanism aims to achieve part or all of the above-mentioned properties while maximizing the overall revenue of buyers $F(x_{l,m,n})$ as given by the following (\ref{e9}), via optimizing the decision matrix $\mathbf{X}=\{x_{l,m,n}|l \in \mathcal{L},m \in \mathcal{M},n \in \mathcal{N}\}$,
\begin{equation}\label{e9}
\begin{aligned}
F(x_{l,m,n})=& \sum \limits_{l \in \mathcal{L}}\sum \limits_{m \in \mathcal{M}} \sum \limits_{n \in \mathcal{N}}x_{l,m,n}v_{l,(m,n)}- \\ & \sum\limits_{l \in \mathcal{L}}\sum \limits_{m \in \mathcal{M}} \sum \limits_{n \in \mathcal{N}}x_{l,m,n}(q_{m,l}+s_{n,(m,l)}).
\end{aligned}
\end{equation}

Thus, the proposed auction is formulated by the following optimization problem, as shown by (\ref{e10}).
{\setlength\abovedisplayskip{3pt}
\setlength\belowdisplayskip{2pt}
\begin{align}
\label{e10}
\text{} \max_{\mathbf{X}}F(x_{l,m,n})
\end{align}
\setcounter{equation}{9}
\begin{subequations}
	\begin{flalign}
	\text{\textit{s.t. ~~~~}} &x_{l,m,n}\in \mathbf{X} \in \{0,1\},  \\
    &\sum_{m \in {\mathcal{M}}} x_{l, m,n} \leq1, \forall l \in \mathcal{L},\\
    & \sum_{l \in \mathcal{L}}x_{l,m,n} \leq 1, \forall m \in \mathcal{M},\\
    &  \sum_{l \in {\mathcal{L}}}\sum_{m \in {\mathcal{M}}}x_{l,m,n}  \leq 1, \forall n \in \mathcal{N},\\
    & d_{m,l}\geq d_{l}, \forall x_{l,m,n}=1, m \in \mathcal{M}, l \in \mathcal{L}.
	\end{flalign}
\end{subequations}
}

Specifically, constraint (10b) ensures that each buyer can only be matched to at most one data-seller, that is, each FL service $l$ can only be assigned to at most one DO. Constraint (10c) indicates each data-seller can only provide service for at most one buyer. Similarly, constraint in (10d) ensures only one UAV-seller can be employed to serve one buyer with a data-seller. Constraint (10e) enforces the winning data-seller should satisfy the data size requirement. Thus problem (\ref{e10}) with constraints (10a)-(10d) presents a 0-1 integer programming problem, which is proved to be NP-Hard and computationally intractable\cite{NP-TNSM,NP-MASS}.

\subsection{Challenges of mechanism design}
Notably, the proposed auction model is difficult to be regarded as a standard reverse auction or double auction due to the following challenges:
\begin{itemize}
  \item \textit{Seller combination}: Different from conventional auctions with single seller type, our proposed auction considers two seller types (data-sellers and UAV-sellers), where the buyer-seller matching problem becomes a complicated tripartite graph matching problem, i.e., a hypergraph matching problem, which is generally NP-Hard\cite{AuctionMatching}. Thus, it is difficult to find efficient algorithms to obtain the optimal solution, especially when facing a large problem size, e.g., large number of buyers and sellers \cite{HyperMatch}. It is worth noting that our proposed auction also differs from the hierarchical or tiered double auctions studied in \cite{HierDA} and \cite{TierDA}, which consider two kinds of sellers selling the same kind of commodities with a superior and subordinate relation; while in the proposed auction, sellers are providing different commodities independently. Besides, our proposed auction applies two different purchase mechanisms for two seller types at the same time, that is, purchasing data set (continuous purchase) from data-sellers, and UAV services (binary purchase) from UAV-sellers, which is different from existing work \cite{New-review3}, that only considers single purchase mechanism at a time.
  \item \textit{Multi-type commodity}: Most existing auctions mainly focus on single item-type, namely, each seller/buyer only sells/buys a single unit of a single type commodity, e.g., McAfee double auction\cite{McAfee}. Although some works have extended the McAfee double auction to multi item-type or multi-unit settings, e.g.,\cite{MIDA-18} and \cite{MUDA-18}, our proposed auction is facing challenges to be solved directly by these methods due to the following reasons. First, each data-seller posses multi-type of commodities (data sets) with a single-unit. Then, each UAV-seller only posses single-type single-unit commodities (UAV services). Also, each buyer only prefers to buy a single item of a specific type from a data-seller. Therefore, our proposed auction model represents a reverse auction with mutually exclusive constraints.
  \item \textit{Heterogeneous bid}: Due to the difference of data volume and unit cost, each data-seller's sell-bids for different data sets are heterogeneous, which lead to a sell-bid vector. Similarly, each UAV-seller also has a heterogeneous bid on data-sellers and buyers, since they jointly determine the cost of each UAV-seller, which requires a sell-bid matrix. Therefore, heterogeneous bids of the participants poses great challenge to solve this problem.
\end{itemize}

\section{VCG-Based Optimal Reverse Auction}
This section introduces an interesting  VCG-based optimal reverse auction mechanism by considering the joint bid associated with data-sellers and UAV-sellers.
\subsection{Seller pair and joint bid}
Each data-seller sells $L$ types of data with single unit, which can thus be regarded as a group of $L$ virtual data-sellers, where each virtual data-seller sells a single data set. Overall, there are $M*L$ virtual data-sellers. Similarly, each UAV-seller has a bid matrix with size $M*L$, and can be decomposed into a set of $M*L$ virtual UAV-sellers, where $N*M*L$ virtual UAV-sellers can be considered. For notational simplicity, we denote the set of $M$ virtual data-sellers selling data set $l$ as $\mathcal{M}_{l}=\{1_{l},\dots,m_{l},\dots,M_{l}\}$, where $m_{l}$ is the virtual data-seller selling data set $l$ of data-seller $m$. Similarly, for any given buyer $l$ and data-seller $m \in \mathcal{M}_{l}$, $N$ feasible virtual UAV-sellers can be denoted as $\mathcal{N}_{l,m}=\{1_{l,m},\dots,n_{l,m},\dots,N_{l,m}\}$.

Accordingly, the proposed auction can be regarded as a reverse auction for each buyer $l$ with $M$ virtual data-sellers and $N*M$ virtual UAV-sellers. Since conventional reverse auction only considers single seller type which brings difficulties in handling the auction procedure, an interesting idea of joint bid is proposed in this paper. The definition of seller pair and joint bid is detailed by the following Definition \ref{Def-JointBidding}.
\begin{myDef}\label{Def-JointBidding}
(Seller pair and joint bid): For each buyer $l \in \mathcal{L}$, the auctioneer combines a data-seller $m \in \mathcal{M}_{l}$ and a UAV-seller $n \in \mathcal{N}_{(l,m)}$ as a seller pair $(m,n)$, and the sum of their individual bid can be termed as a joint bid $J_{l,(m,n)}$, i.e., $J_{l,(m,n)}=q_{m,l}+s_{n,m,l}$.
\end{myDef}

Since the sell-bids of each $\mathcal{M}_{l}$ represents a vector of size $1*M$ and is a matrix of size $N*M$ of $\mathcal{N}_{(l,m)}$, we can use a matrix $J_{l,(m,n)}\in \mathbf{J}_{l}$ of size $N*M$ to represent the joint bid of seller pairs regarding each buyer. Accordingly, the auctioneer can conduct the proposed reverse auction based on joint bid.


\begin{algorithm}[t]\label{Alg-1}
    \small
	\caption{VCG-Based Optimal Reverse Auction Algorithm}
	\begin{algorithmic}[1]
    \Require Buyer set $\mathcal{L}$, seller sets $\mathcal{M}$, $\mathcal{N}$, bid information of each seller and buyer $B^{d}_{m}$, $\vec{c}_{m}$, $\mathbf{E}_{n}$, $B_{n}^{u}$, $d_{l}$, $\mathbf{V}_{l}$.
    \Ensure Winner determination result $\mathbf{X}$, final payment profile $\vec{P}^{d}$ of data-sellers and $\vec{P}^{u}$ of UAV-sellers.
    \State Initializes $\mathbf{X}$ with all zeros. Virtualize data-sellers, UAV-sellers, and obtain the joint bid matrix $\mathbf{J}_{l}$ for each buyer $l$.
	\State Solve the problem (\ref{e10}) and obtain the optimal solution (winning buyers and sellers) $\mathbf{X}^{*}$.
    \For {any buyer-seller pair $(l,i,k)$}
        \If{$x_{l,i,k}=1$}
        \State Calculate the total payment $P^{f}_{i,k}$ to the winning seller pair $(i,k)$ with the following payment rule defined in (\ref{e11}).
        \State Calculate the received payment of the winning data-seller $i$ and UAV-seller $k$ from any winning buyer $l$ as:
        \Statex $\qquad\quad p_{i}^{d}=\frac{q_{i,l}}{J_{l,(i,k)}}P^{f}_{i,k}$, $p_{k}^{u}=\frac{s_{k,i,l}}{J_{l,(i,k)}}P^{f}_{i,k}$.
         \Else
         \State $p_{i}^{d}\Leftarrow 0$, $p_{k}^{u}\Leftarrow 0$.
          \EndIf
    \EndFor
	\end{algorithmic}
\end{algorithm}
\subsection{VCG-based optimal reverse auction algorithm}
The proposed VCG-based optimal reverse auction for Multiple FL service market is summarized in \textbf{Algorithm 1}. Specifically, we can obtain the joint bid of each seller pair for each buyer by line 1. Getting the optimal solution of problem (\ref{e10}) relies on determining the winning buyers and seller-pairs (line 2 in Algorithm 1), which represents the key issue to ensure the truthfulness of VCG mechanism. Moreover, VCG-based payment rule determines the final total payment $P_{i,k}^{f}$ of the winning seller pair $(i,k)$ with joint bid $J_{l,(i,k)}$ as follows,
\begin{equation}\label{e11}
P^{f}_{i,k}=F(x_{l,m,n}^{*})-F_{\setminus(i,k)}(y_{l,m,n}^{*})+ J_{l,(i,k)},
\end{equation}
where $F(x_{l,m,n}^{*})$ is the the value of function (\ref{e9}) under the optimal decision matrix $\mathbf{X}^{*}$ and $F_{\setminus(i,k)}(y_{l,m,n}^{*})$ represents the optimal result when data-seller $i$ and UAV-seller $k$ do not join in the auction, and $y_{l,m,n}^{*}$ stands for the optimal solution without seller pair $(i,k)$. From lines 6-9 in Algorithm 1, the final payment can be allocated to the data-seller and UAV-seller in proportion to their sell bids.
\subsection{Complexity and properties analysis}
The computational complexity of Algorithm 1 is dominated by the optimal solution of the problem (\ref{e10}). The computation complexity of winner determination is $\mathcal{O}(2^{LMN})$ and $\mathcal{O}(L2^{LMN})$ for payment rule.

Since the joint bid is the sum of sell-bids of both data-seller and UAV-seller, the auction properties can be proved under considering the following three cases:
\begin{itemize}
  \item \textit{Case 1}: Both sellers bid truthfully and the joint bid equals to the total cost.
  \item \textit{Case 2}: At least one seller bids untruthfully and the joint bid does not equal to the total cost.
  \item \textit{Case 3}: Both sellers bid untruthfully but the joint bid still equals to the total cost, e.g., one underbids and the other overbids.
\end{itemize}
Then, we first focus on the first two cases and prove the overall truthfulness of seller pair. Based on definitions in Section III, we give the definition of seller pair's revenue as,
\begin{myDef}\label{Def-SP}
(Revenue of seller pair): For any seller pair $(i,k)$ associated with data-seller $i \in \mathcal{M}_{l}$ and UAV-seller $n \in \mathcal{N}_{l,i}$, the corresponding revenue is defined as the difference between the final total payment $P_{i,k}^{f}$ and the true joint bid $J_{l,(i,k)}$,
\begin{equation}\label{e12}
U_{(i,k)}= \sum \limits_{l \in \mathcal{L}}x_{l,i,k}\left(P_{i,k}^{f}-J_{l,(i,k)}\right).
\end{equation}
\end{myDef}

\begin{myTheo}
(Truthfulness of seller pair): In the VCG-based reverse auction of FL service market, for any seller pair $(i,k)$ where data-seller $i \in \mathcal{M}_{l}$ and UAV-seller $k \in \mathcal{N}_{l,i}$, having joint bid equals to the total cost (i.e., $J_{l,(i,k)}=c_{i,l}+e_{k,(i,l)}$) is a weakly dominant strategy.
\end{myTheo}

\begin{proof}
Based on the Definition \ref{Def-SP} and (\ref{e11}), for any given winning buyer $l$ and seller pair $(i,k)$, the seller pair's revenue can be denoted as follows with truthful bid,
\begin{equation}\label{e13}
U_{(i,k)}=F(x_{l,m,n}^{*})-F_{\setminus(i,k)}(y_{l,m,n}^{*}),
\end{equation}\
and the revenue when seller pair $(i,k)$ bids untruthfully with $\bar{J}_{l,(i,k)}$ is
\begin{equation}\label{e14}
\begin{aligned}
\overline{U}_{(i,k)}= & {F}(\overline{x}_{l,m,n}^{*})-F_{\setminus(i,k)}(\overline{y}_{l,m,n}^{*})+\overline{J}_{l,(i,k)}-
\\
& (q_{i,l}+s_{k,(i,l)}).
\end{aligned}
\end{equation}

Note that $F_{\setminus(i,k)}(y_{l,m,n}^{*})=F_{\setminus(i,k)}(\overline{y}_{l,m,n}^{*})$, since they both denote the optimal solution excluding seller pair $(i,k)$, the difference of $U_{(i,k)}$ and $\overline{U}_{(i,k)}$ can be denoted as
\begin{equation}\label{e15}
\begin{aligned}
U_{(i,k)}-\overline{U}_{(i,k)}= &F(x_{l,m,n}^{*})-{F}(\overline{x}_{l,m,n}^{*})-\overline{J}_{l,(i,k)}+
\\   & (q_{i,l}+s_{k,(i,l)}),
\end{aligned}
\end{equation}
where
\begin{equation}\label{e16}
\begin{aligned}
F(x_{l,m,n}^{*})= & \sum \limits_{l \in \mathcal{L}}\sum \limits_{m \in \mathcal{M}_{l}} \sum \limits_{n \in \mathcal{N}_{l,m}}x_{l,m,n}^{*}v_{l,(m,n)}- \\ & \sum\limits_{l \in \mathcal{L}}\sum \limits_{m \in \mathcal{M}_{l}} \sum \limits_{n \in \mathcal{N}_{l,m}}x_{l,m,n}^{*}J_{l,(m,n)},
\end{aligned}
\end{equation}
and
\begin{equation}\label{e17}
\begin{aligned}
F(\overline{x}_{l,m,n}^{*})=& \sum \limits_{l \in \mathcal{L}}\sum \limits_{m \in \mathcal{M}_{l}} \sum \limits_{n \in \mathcal{N}_{l,m}}\overline{x}_{l,m,n}^{*}v_{l,(m,n)}- \\ & \sum\limits_{l \in \mathcal{L}}\sum \limits_{m \in \mathcal{M}_{l}, \atop m \neq i} \sum \limits_{n \in \mathcal{N}_{l,m},\atop n \neq k}\overline{x}_{l,m,n}^{*}J_{l,(m,n)}-
\\
&\overline{J}_{l,(i,k)}.
\end{aligned}
\end{equation}

By substituting (\ref{e16}) and (\ref{e17}) into (\ref{e15}), we have
\begin{equation}\label{e18}
\begin{aligned}
U_{(i,k)}-\overline{U}_{(i,k)}
&= \left( {\sum \limits_{l \in \mathcal{L}}\sum \limits_{m \in \mathcal{M}_{l}} \sum \limits_{n \in \mathcal{N}_{l,m}}x_{l,m,n}^{*}v_{l,(m,n)}}- \right. \\
& \left.{ \sum\limits_{l \in \mathcal{L}}\sum \limits_{m \in \mathcal{M}_{l}} \sum \limits_{n \in \mathcal{N}_{l,m}}x_{l,m,n}^{*}J_{l,(m,n)}} \right)-\\
& \left( { \sum \limits_{l \in \mathcal{L}}\sum \limits_{m \in \mathcal{M}_{l}} \sum \limits_{n \in \mathcal{N}_{l,m}}\overline{x}_{l,m,n}^{*}v_{l,(m,n)}}- \right. \\
& \left. {\sum\limits_{l \in \mathcal{L}}\sum \limits_{m \in \mathcal{M}_{l}} \sum \limits_{n \in \mathcal{N}_{l,m}}\overline{x}_{l,m,n}^{*}J_{l,(m,n)}} \right)
\end{aligned}
\end{equation}

Since $x_{l,m,n}^{*}$ is the optimal solution of (\ref{e10}), which can definitely achieve an overall revenue larger than or equal to other solutions (e.g, $\overline{x}_{l,m,n}^{*}$). Thus, $U_{(i,k)}-\overline{U}_{(i,k)}\geq 0$ always holds according to (\ref{e18}), which indicates each seller pair has no incentive to bid untruthfully.
\end{proof}
\begin{myTheo}
(Individual rationality of seller pair): In the VCG-based reverse auction of FL service market, for any data-seller $i \in \mathcal{M}_{l}$ and UAV-seller $k \in \mathcal{N}_{l,i}$, seller pair $(i,k)$ possesses the property of individual rationality.
\end{myTheo}
\begin{proof}
Given that all seller pairs are truthful in VCG-based reverse auction, the revenue of any seller pair $(i,k)$ can be calculated by (\ref{e13}). Since $F(x_{l,m,n}^{*})$ stands for the optimal solution of upon considering all the participants and $F_{\setminus(i,k)}(y_{l,m,n}^{*})$ represents the optimal solution excluding seller pair $(i,k)$, the solution space of $y_{l,m,n}^{*}$ can be regarded as a subset of that of $x_{l,m,n}^{*}$. Thus, we have $F(x_{l,m,n}^{*})-F_{\setminus(i,k)}(y_{l,m,n}^{*})\geq 0$. Consequently, each seller pair is individual rational and can obtain non-negative revenue when winning the auction.
\end{proof}

\begin{myCoro}
In the VCG-based reverse auction of FL service market, each data-seller $i \in \mathcal{M}_{l}$ and UAV-seller $k \in \mathcal{N}_{l,i}$ of seller pair $(i,k)$ is individual rational.
\end{myCoro}
\begin{proof}
The above-mentioned analysis has proved that each seller pair is individual rational. According to the payment rule in Algorithm 1, it is straight to conclude that each data-seller or UAV-seller can obtain non-negative revenue when winning the auction. Thus each individual seller of each seller pair is individual rational.
\end{proof}

\begin{myCoro}
In the VCG-based reverse auction of FL service market, for any data-seller $i \in \mathcal{M}_{l}$ and UAV-seller $k \in \mathcal{N}_{l,i}$, having sell-bids equals to the cost ($q_{i,l}=c_{i,l},s _{k,(i,l)}=e_{k,(i,l)}$) is a weakly dominant strategy.
\end{myCoro}

\begin{proof}
Although we have proved that having the joint bid equals to the total cost is a weakly dominant strategy for any seller pairs, \textit{Case 3} may exist when considering both the sellers bid untruthfully, while the joint bid still equals to the total cost. However, since we assume that there is no collusion or negotiation between the sellers, no individual seller has incentives to bid untruthfully. This can be proved via a simple example.

Suppose that an untruthful seller pair $(i,k)$, with the corresponding joint bid $J_{l,(i,k)}$ equals to the total cost, where data-seller $i$ underbids with $\overline{q}_{i,l}$ and UAV-seller $k$ overbids with $\overline{s}_{k,(i,l)}$ where $\overline{q}_{i,l}+\overline{s}_{k,(i,l)}=J_{l,(i,k)}$, $\overline{q}_{i,l}=q_{i,l}-\sigma$, $\overline{s}_{k,(i,l)}=q_{i,l}+\sigma$, and $\sigma$ is positive. Let the seller pair's total payment with truthful joint bid $J_{l,(i,k)}$ be $P^{f}_{i,k}$. Then, the data-seller's revenue with truthful and untruthful bid can be denoted as (\ref{e19}), and (\ref{e20}), respectively,
\begin{equation}\label{e19}
U_{i}^{d}=\frac{q_{i,l}}{J_{l,(i,k)}}P^{f}_{i,k}-q_{i,l},
\end{equation}
\begin{equation}\label{e20}
\overline{U}_{i}^{d}=\frac{\overline{q}_{i,l}}{J_{l,(i,k)}}P^{f}_{i,k}-\overline{q}_{i,l}.
\end{equation}

The difference of $U_{i}^{d}$ and $\overline{U}_{i}^{d}$ can be denoted as
\begin{equation}\label{e21}
\begin{aligned}
U_{i}^{d}-\overline{U}_{i}^{d}&= \frac{P^{f}_{i,k}}{J_{l,(i,k)}}(q_{i,l}-\overline{q}_{i,l})-(q_{i,l}-\overline{q}_{i,l}) \\
&= \left(\frac{P^{f}_{i,k}}{J_{l,(i,k)}}-1\right)(q_{i,l}-\overline{q}_{i,l}).
\end{aligned}
\end{equation}

As $(q_{i,l}-\overline{q}_{i,l})=\sigma >0$, and $\frac{P^{f}_{i,k}}{j_{l,(i,k)}}\geq1$ holds based on the individual rationality of seller pair. Thus, $U_{i}^{d}-\overline{U}_{i}^{d}\geq 0$ always holds, which indicates sellers cannot obtain larger revenue by underbidding. On the contrary, sellers have incentives to overbids as long as remain the joint bid truthful. However, since we assume there is no collusion and interest transfer between the sellers, thus no sellers have incentives to sacrifice their revenue to subsidize the others. Thus, truthful bidding is a weakly dominant strategy for individual seller in each seller pair.

\end{proof}

Finally, we can conclude that the proposed VCG-based reverse auction of FL service market is strategy-proof, which, however, is not computationally efficient.

\section{One-Sided Matching-Based Suboptimal Reverse Auction}
Although the VCG-based reverse auction can ensure the optimality, as well as the truthfulness and individual rationality of sellers, the exponential complexity of winner determination and payment rule impedes it in practical application with a large number of auction participants. Thus, this section proposes a computationally efficient winner determination and payment rule based on one-sided matching, which can also ensure the truthfulness and individual rationality of sellers.

\subsection{One-sided matching-based mechanism design}
To maximize the overall revenue of buyers, the auctioneer will match the buyers to feasible sellers with maximum overall revenue, while each buyer prefers to select sellers to maximize its revenue as well. Moreover, sellers sell their commodities to buyers to obtain revenue but without initiatives in the reverse auction. Thus, preference priority can be considered from the perspective of auctioneer and buyers in such a one-to-one one-sided matching. Based on (\ref{e8}), we define the preference value for any buyer-seller pair $(l,m,n), l \in \mathcal{L}, m \in \mathcal{M}_{l}, n \in \mathcal{N}_{l,m}$ as:
\begin{equation}\label{e22}
R_{l,m,n}=v_{l,(m,n)}-J_{l,(m,n)}.
\end{equation}

From the view of auctioneer, it can hold a preference list denoted by $\mathbf{T_{A}}$ (which contains the indexes of buyer-seller pairs) for any buyer-seller pair in a non-ascending order so as to maximize the overall revenue. Accordingly, the preference list can be established based on the preference relationship as (\ref{e23}),
\begin{equation}\label{e23}
\mathbf{T_{A}}: (l,m,n)\succ_{A} (l,i,k)\Leftrightarrow R_{l,m,n}>R_{l,i,k},
\end{equation}
where $\succ_{A}$ means the auctioneer prefers the left than the right.

Similarly, each buyer $l$ can establish a preference list $\mathbf{T_{l}}$ (indexes of seller pairs) by sorting the preference value $R_{l,m,n},m \in \mathcal{M}_{l}, n \in \mathcal{N}_{l,m}$ in a non-ascending order, which can be denoted as,
\begin{equation}\label{e24}
\mathbf{T_{l}}: (l,(m,n))\succ_{l} (l,(i,k))\Leftrightarrow R_{l,m,n}>R_{l,i,k}.
\end{equation}

Finally, we can obtain $\mathbf{T_{A}}$ and $\mathbf{T_{l}}$ for both the auctioneer and each buyer $l$, respectively. To ensure the truthfulness of the sellers, we add a virtual \textit{NULL seller pair} at the end of $\mathbf{T_{l}}$ with preference value equals 0, which describing that if no real sellers wins the auction for the buyer, the buyer is assigned to a null seller pair. Particularly, if any preference value $R_{l,m,n}<0$, then buyer-seller pair $(l,m,n)$ will not be stored in $\mathbf{T_{A}}$ and $\mathbf{T_{l}}$ to prevent non-positive revenue.

Obviously, each buyer tends to select the top sell pairs in its preference list. However, as each seller pair can only trade with one buyer, the auctioneer prefers to assign the seller pair to the buyer with higher preference value. For example, assume seller pair $(m,n)$ is on the top of list of buyer $l$ and $l^{\prime}$ at the same time, the auctioneer assigns seller pair $(m,n)$ to buyer $l$ if $R_{l,m,n}>R_{l^{\prime},m,n}$. Considering the preference lists of the auctioneer and buyers, winning buyer-seller pairs are matched accordingly in the one-sided matching process until all buyers are matched.

\subsection{Suboptimal winner determination and payment rule}
To facilitate the presentation of our proposed winner determination and payment rule, the definition of \textit{critical value} is given below:
\begin{myDef}
(Critical value): The critical value of any buyer-seller pair $(l,m,n)$ in one-sided matching is defined as
\begin{equation}\label{e25}
\widetilde{R_{l,m,n}}=v_{l,\widetilde{(m,n)}}-J_{l,\widetilde{(m,n)}},
\end{equation}
where $\widetilde{(m,n)}$ is the first seller pair located behind seller pair $(m,n)$ in the preference list $\mathbf{T_{l}}$.
\end{myDef}

The overall one-sided matching-based suboptimal reverse auction mechanism mainly compromises two phases: 1) one-sided one-to-one matching based suboptimal winner determination; 2) and payment calculation. We first give the payment rule based on the one-sided matching result. Specifically, the total payment $P_{i,k}^{f}$ of the winning seller pair $(i,k)$ from buyer $l$ is calculated as
\begin{equation}\label{e26}
P_{i,k}^{f}=v_{l,(i,k)}-\widetilde{R_{l,i,k}}.
\end{equation}

We summarize the one-sided matching based suboptimal reverse auction in \textbf{Algorithm 2}. By applying lines 1-4, we calculate the preference value by given the input parameter, then we build the preference list for the auctioneer and buyers by sorting the preference value in non-ascending order in lines 5-6. Lines 8-14 illustrate the one-sided matching process. Finally, the final payment of the winning seller is calculated in lines 15-21.

\begin{algorithm}[t]\label{Alg-2}
\small
	\caption{One-Sided Matching-Based Suboptimal Reverse Auction Algorithm}
	\begin{algorithmic}[1]
    \Require Buyer set $\mathcal{L}$, seller sets $\mathcal{M}$, $\mathcal{N}$, bid information of each seller and buyer $B^{d}_{m}$, $\vec{c}_{m}$, $\mathbf{E}_{n}$, $B_{n}^{u}$, $d_{l}$, $\mathbf{V}_{l}$.
    \Ensure Winner determination result $\mathbf{X}$, final payment profile $\vec{P}^{d}$ of data-sellers and $\vec{P}^{u}$ of UAV-sellers.
    \State Initializes $\mathbf{X}$ with all zeros.Virtualize data-sellers, UAV-sellers, and obtain the joint matrix $\mathbf{J}_{l}$ for each buyer $l$.
	 \For { buyer $l \in \mathcal{M}$ and seller pair $(i,k), i \in \mathcal{M}_{l}, k \in \mathcal{N}_{l,i}$}
      \State Calculate the preference value $R_{l,i,k}$ according to (\ref{e22}).
     \EndFor
     \State $\mathbf{T_{A}} \Leftarrow Sort(l \in \mathcal{L}, i \in \mathcal{M}_{l}, k \in \mathcal{N}_{l,i}| R_{l,i,k}, ``non-ascending'')$.
     \State $\mathbf{T_{l}} \Leftarrow Sort(i \in \mathcal{M}_{l}, k \in \mathcal{N}_{l,i}| R_{l,i,k}, ``non-ascending'')$.
     \State $\mathcal{W}\Leftarrow \mathcal{M}$.

     \While {$\mathbf{T_{A}}\neq \varnothing$ and $\mathcal{W} \neq \varnothing$} \%\textit{Phase 1:  Winner determination}
        \For{each first entry $(l,i,k) \in \mathbf{T_{A}}$}
            \State $x_{l,i,k}\Leftarrow1$
            \State $\mathbf{T_{A}}\Leftarrow \mathbf{T_{A}}\setminus$\{all elements related to $l,i,k$\}
            \State $\mathcal{W}\Leftarrow\mathcal{W}\setminus\{l\}$
        \EndFor
     \EndWhile
     \For {$x_{l,i,k}\in \mathbf{X}$} \%\textit{Phase 2: Payment calculation}
        \If{$x_{l,i,k}=1$}
             \State Calculate the total payment $P^{f}_{i,k}$ of seller pair $(i,k)$ according to (27).
             \State Calculate the received payment of the winning data-seller $i$ and UAV-seller $k$ as:
             \Statex $\qquad\quad p_{i}^{d}=\frac{q_{i,l}}{J_{l,(i,k)}}P^{f}_{i,k}$, $p_{k}^{u}=\frac{s_{k,i,l}}{J_{l,(i,k)}}P^{f}_{i,k}$.
             \Else
             \State $p_{i}^{d}\Leftarrow 0$, $p_{k}^{u}\Leftarrow 0$.
        \EndIf
      \EndFor

%
	\end{algorithmic}
\end{algorithm}

\subsection{Complexity and properties analysis}
In this subsection, we analyze the complexity and properties of  the proposed one-sided matching-based reverse auction. Moreover, we prove that the one-sided matching-based reverse auction is stable.

\begin{myTheo}
(Computational efficiency): The proposed one-sided matching-based reverse auction algorithm is computationally efficient.
\end{myTheo}
\begin{proof}
In Algorithm 2, the time complexity of line 1-4 is $\mathcal{O}(LMN)$ and the two sort procedures in line 5 and 6 is $\mathcal{O}(LMNlog(LMN))$ and $\mathcal{O}(MNlog(MN))$, respectively \cite{STAR}. Similarly, the time complexity of line 8-21 is $\mathcal{O}(2LMN)$. Thus, Algorithm 2 has overall polynomial time complexity.
\end{proof}

\begin{myTheo}
(Truthfulness of seller pair): The proposed one-sided matching-based reverse auction of FL service market is truthful for each seller pair.
\end{myTheo}
\begin{proof}
First, for any given seller pair $(i,k)$ with truthful bid $J_{l,(i,k)}=c_{i,l}+e_{k,(i,l)}$, according to (\ref{e12}), (\ref{e25}) and (\ref{e26}), the seller pair's revenue can be denoted as,
\begin{equation}\label{e27}
U_{(i,k)}=v_{l,(i,k)}-\widetilde{R_{l,i,k}}-J_{l,(i,k)}.
\end{equation}

Meanwhile, we assume a virtual seller pair $(i^{\prime},k^{\prime})$ who is exactly the same as the seller pair $(i,k)$ except the sell-bid, is considered as the case that seller pair $(i,k)$ bids untruthfully with a different joint bid ${J_{l,(i,k)}}^{\prime}$, thus its revenue can be denoted as,
\begin{equation}\label{e28}
U_{(i,k)}^{\prime}= {v_{l,(i,k)}}^{\prime}-\widetilde{R_{l,i,k}}^{\prime}-{J_{l,(i,k)}}.
\end{equation}

Then, $U_{(i,k)}-U_{(i,k)}^{\prime}$ can be denoted as,
\begin{equation}\label{e29}
\begin{aligned}
U_{(i,k)}-U_{(i,k)}^{\prime}& =  v_{l,(i,k)}-\widetilde{R_{l,i,k}}-{v_{l,(i,k)}}^{\prime}+\widetilde{R_{l,i,k}}^{\prime} \\
& = \widetilde{R_{l,i,k}}^{\prime}-\widetilde{R_{l,i,k}},
\end{aligned}
\end{equation}
note that $v_{l,(i,k)}={v_{l,(i,k)}}^{\prime}$ since they have same properties.

Generally, there are two cases: seller pair $(i,k)$ wins or loses in the original auction. First, we assume that seller pair $(i,k)$ used to be a winner and obtains revenue $U_{(i,k)}$.
\begin{itemize}
  \item[1)] If ${J_{l,(i,k)}}^{\prime}<J_{l,(i,k)}$, in this case, seller pair $(i^{\prime},k^{\prime})$ always wins as it must be located before $(i,k)$ or at its original position in $\mathbf{T_{l}}$, which means $\widetilde{R_{l,i,k}}^{\prime}\geq\widetilde{R_{l,i,k}}$, thus we have $U_{(i,k)}-U_{(i,k)}^{\prime}\geq0$.
  \item[2)] If ${J_{l,(i,k)}}^{\prime}>J_{l,(i,k)}$, it is obvious that seller pair $(i^{\prime},k^{\prime})$ will locate at the same position or behind $(i,k)$ in $\mathbf{T_{l}}$: if $(i^{\prime},k^{\prime})$ stays at the original position, then it obtains the revenue $U_{(i,k)}^{\prime}=U_{(i,k)}$ according to the payment rule; if seller pair $(i^{\prime},k^{\prime})$ located behind $(i,k)$ in $\mathbf{T_{l}}$, then $(i^{\prime},k^{\prime})$ definitely loses and gets revenue $U_{(i,k)}^{\prime}=0$, as there must be another seller pair can offer higher revenue to be matched with buyer $l$.
\end{itemize}

Second, we consider seller pair $(i,k)$ used to be a loser and obtains $U_{(i,k)}=0$ :
\begin{itemize}
  \item[1)] If ${J_{l,(i,k)}}^{\prime}<J_{l,(i,k)}$, in this case, there are possibilities that seller pair $(i^{\prime},k^{\prime})$ locates at the same position or before $(i,k)$. Specifically, if $(i^{\prime},k^{\prime})$ stays at the original location, then it still loses and gets revenue $U_{(i,k)}^{\prime}=0$. If $(i^{\prime},k^{\prime})$ locates before $(i,k)$ and loses, it still obtains revenue $U_{(i,k)}^{\prime}=0$; if $(i^{\prime},k^{\prime})$ locates before $(i,k)$ and wins, then it gets revenue $U_{(i,k)}^{\prime}$, according to (\ref{e22}) and (\ref{e28}), $U_{(i,k)}^{\prime}$ can be rewrite as follows:
    \begin{equation}\label{e30}
     \begin{aligned}
    U_{(i,k)}^{\prime}=R_{l,i,k}- \widetilde{R_{l,i,k}}^{\prime},
    \end{aligned}
    \end{equation}
    thus if seller pair $(i^{\prime},k^{\prime})$ is in front of $(i,k)$, we have $R_{l,i,k}<\widetilde{R_{l,i,k}}^{\prime}$ and $U_{(i,k)}^{\prime}<U_{(i,k)}=0$, which means it is not encouraged for an individual rational seller.
  \item[2)] If ${J_{l,(i,k)}}^{\prime}>J_{l,(i,k)}$, obviously, seller pair $(i^{\prime},k^{\prime})$ always loses and obtains revenue $U_{(i,k)}^{\prime}=0$.
\end{itemize}

Thus, we can conclude that any seller pair $(i,k)$ cannot obtain higher revenue by bidding untruthfully.
\end{proof}

\begin{table*}[t]
\centering
\begin{center}\renewcommand\arraystretch{1.2}
     \centering
     \caption{Running time (seconds) of different methods considering various number of buyers/UAV-sellers/data-sellers}
     \label{tab2}
     \setlength{\tabcolsep}{1.0mm}{
\begin{tabular}{|c|c|c|c|c|c|c|c|c|c|}
\hline
\diagbox{Methods}{Problem size}&
1/5/5&3/5/5&5/5/5&7/5/5&9/5/5&1/3/5&2/4/6&3/5/7&4/6/8 \\
\hline
Opt& 10097 & 11384 & 13771 & 27799 & 2213192 & 7478 & 11901 & 25048 & 151866 \\
\hline
Subopt& 0.358 & 0.558 & 1.304& 1.097& 1.303 & 0.266&0.432 & 0.743 & 1.093\\
\hline
FOGA& 3.94 & 5.43 & 15.50 & 81.01& 1137.99& 0.73&5.25& 192.16 & 17870.20\\
\hline
HVPM& 0.305 & 0.321 & 0.388 & 0.414& 0.421& 0.250&0.309& 0.321& 0.364\\
\hline
LCPM&  0.277& 0.331& 0.350 & 0.411 & 0.424 & 0.279 &0.310& 0.329 & 0.382 \\
\hline
RSBM&  0.305& 0.400& 0.407 & 0.427 & 0.434 & 0.292 &0.336& 0.342 & 0.354 \\
\hline
\end{tabular}}

\end{center}
\end{table*}

\begin{figure*}[t]
    \centering
    \subfigure[] {\includegraphics[width=1.7in,angle=0]{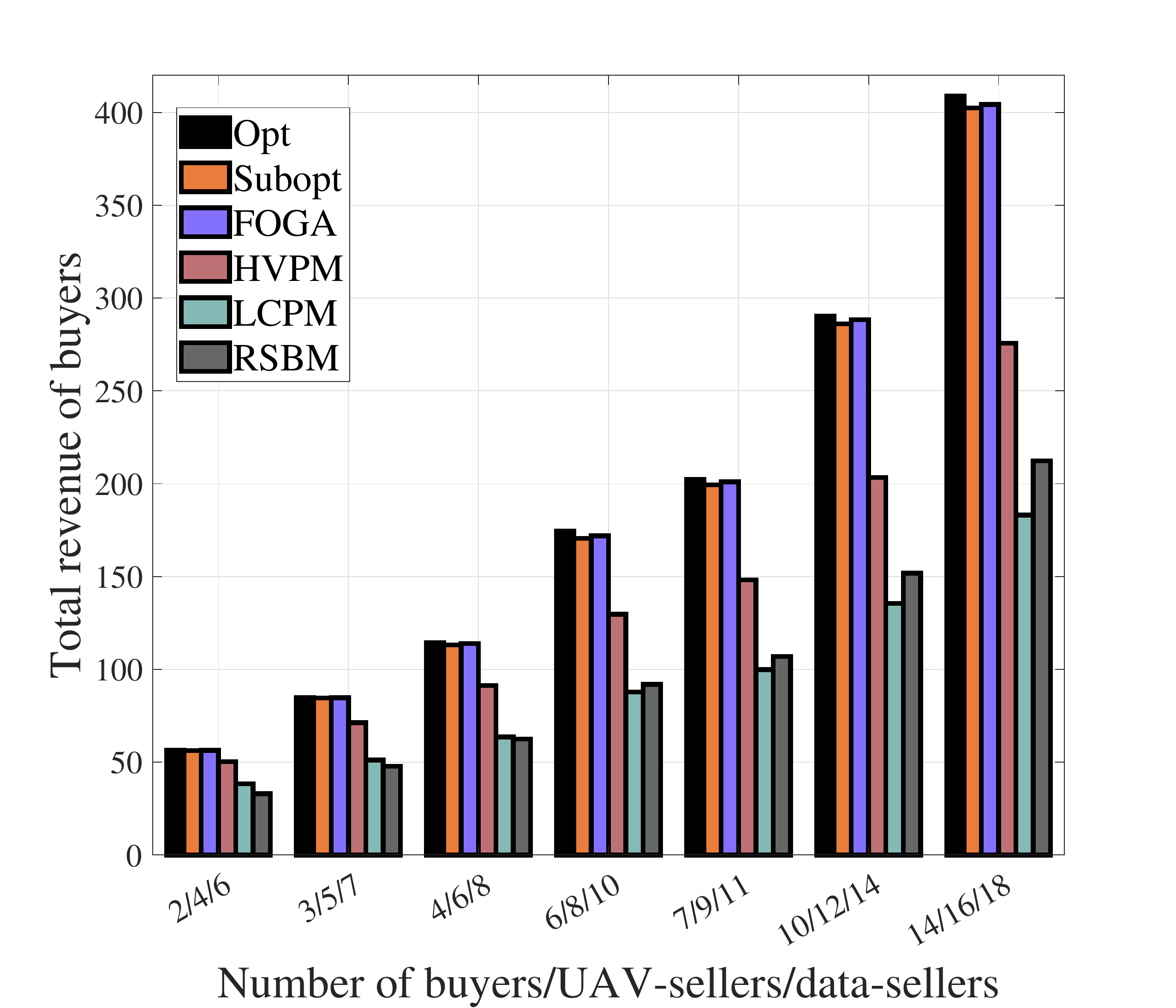}}
    \subfigure[] {\includegraphics[width=1.7in,angle=0]{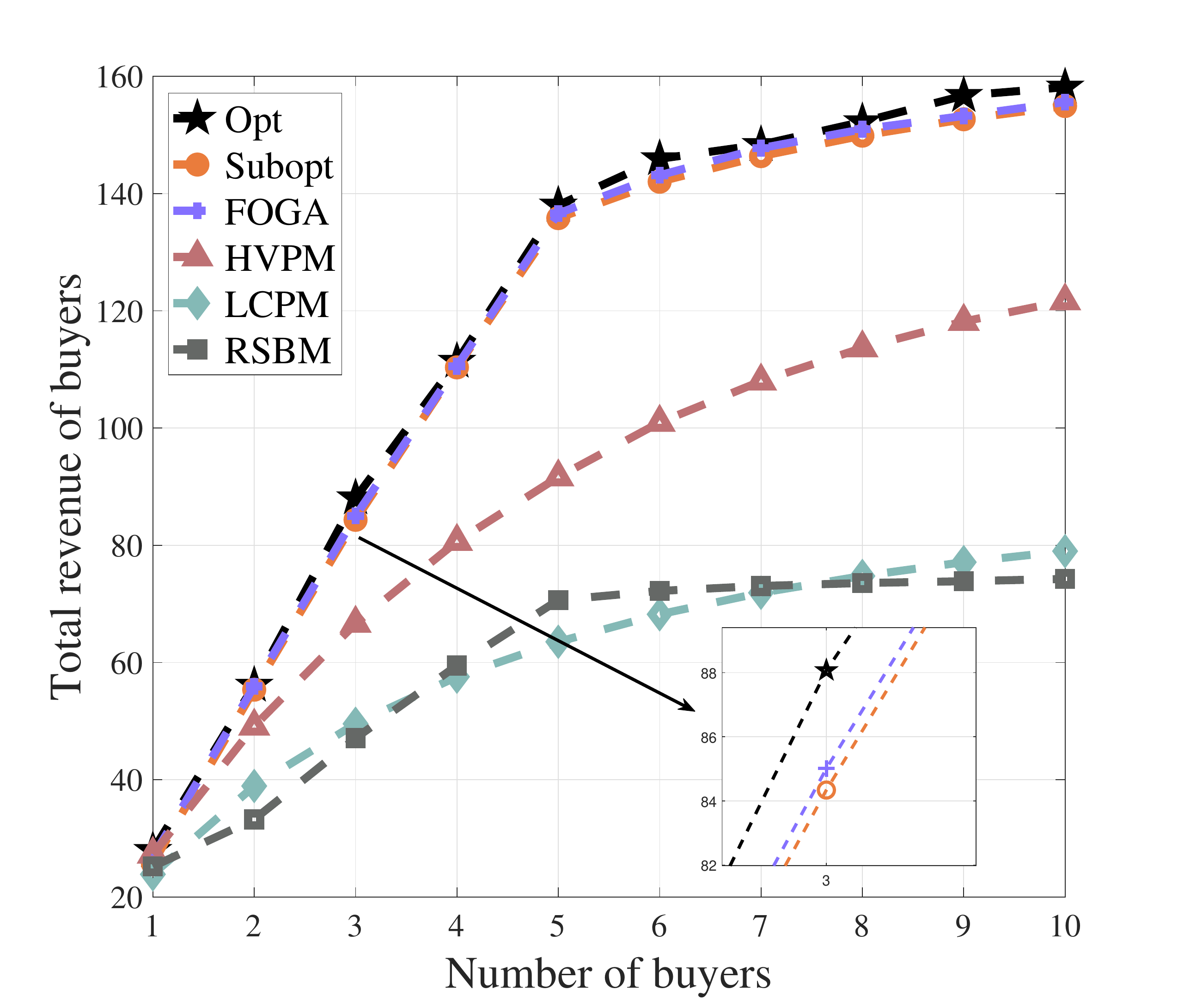}}
    \subfigure[] {\includegraphics[width=1.7in,angle=0]{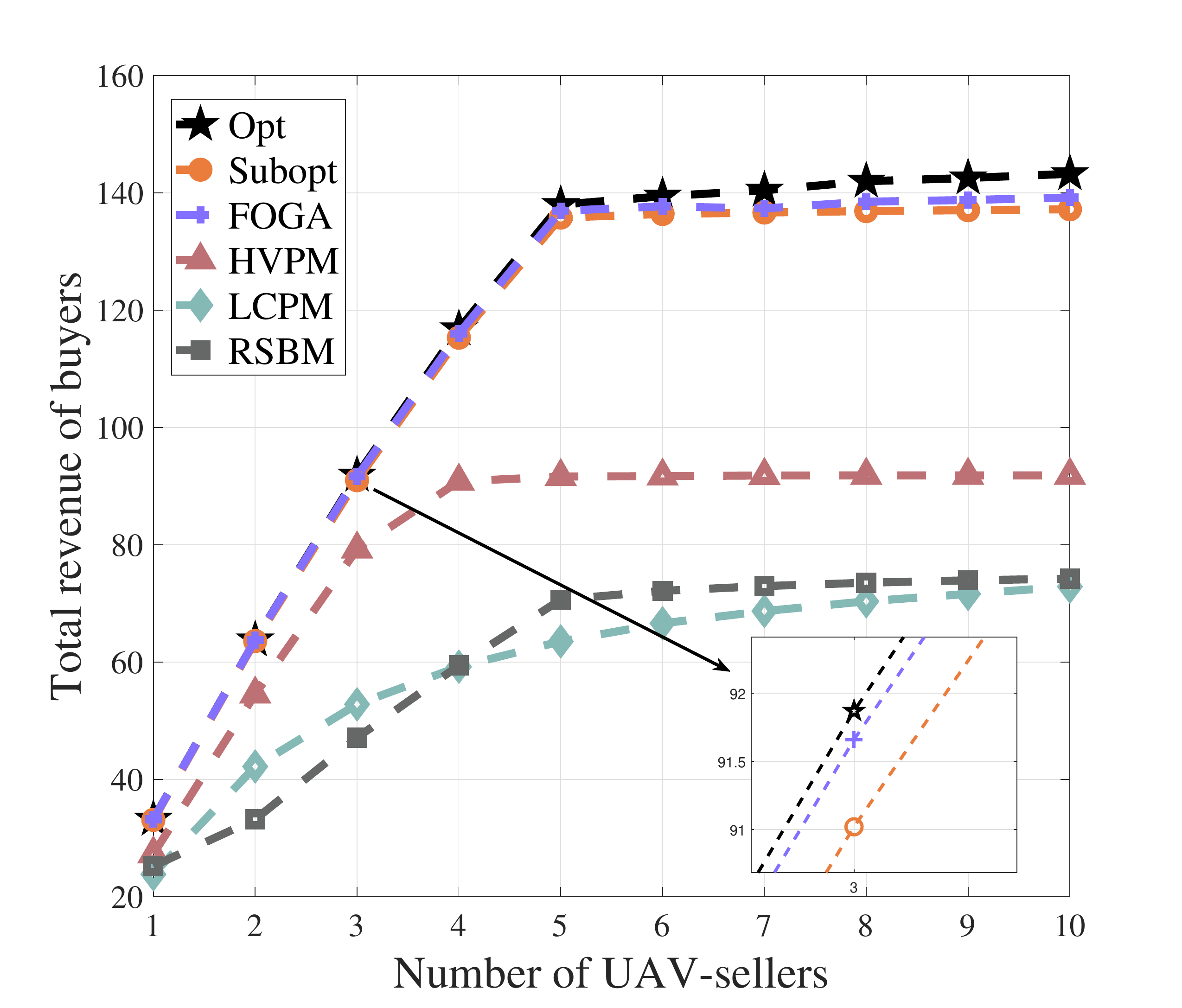}}
    \subfigure[] {\includegraphics[width=1.7in,angle=0]{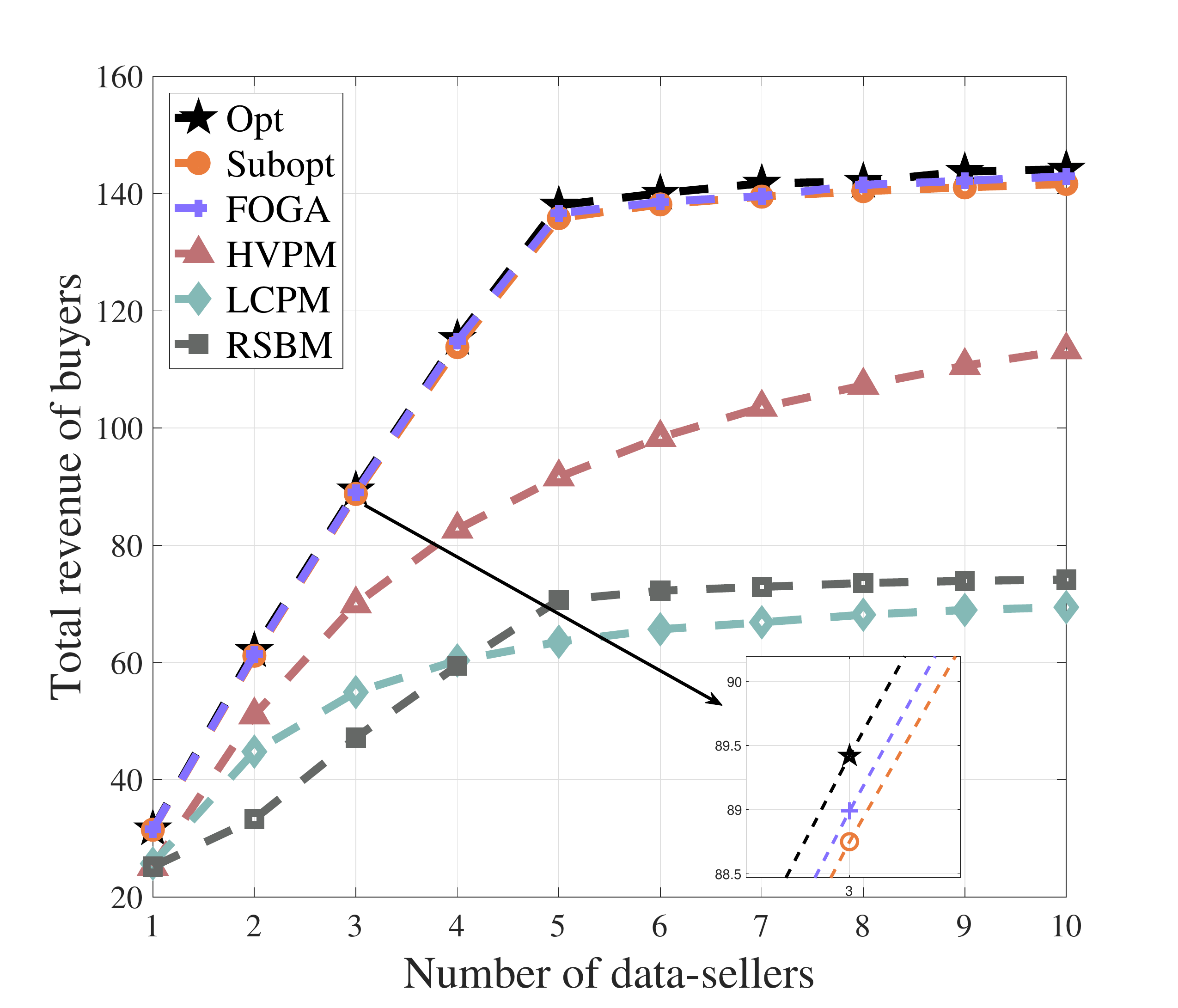}}
   \caption{The obtained total revenue of buyers of different methods versus the number of buyers/UAV-sellers/data-sellers.}
    \label{fig3}
\end{figure*}

\begin{myTheo}
(Individual rationality of seller pair): The proposed one-sided matching-based reverse auction of FL service market is individually rational for each seller pair.
\end{myTheo}
\begin{proof}
As denoted in (\ref{e27}), the revenue of any winning seller pair $(i,k)$ can be further denoted as,
\begin{equation}\label{e31}
\begin{aligned}
U_{(i,k)}& =v_{l,(i,k)}-\widetilde{R_{l,i,k}}-J_{l,(i,k)}\\
& = (v_{l,(i,k)}-J_{l,(i,k)})-\widetilde{R_{l,i,k}}\\
& = R_{l,i,k}- \widetilde{R_{l,i,k}}.
\end{aligned}
\end{equation}
Since $\widetilde{R_{l,i,k}}$ is the preference value of seller pair located behind $(i,k)$ in $\mathbf{T_{l}}$, thus $R_{l,i,k}- \widetilde{R_{l,i,k}}\geq0$ always holds, thus each seller pair can obtain non-negative revenue when it wins the auction.
\end{proof}

Similarly, based on the truthfulness and individual rationality of seller pair, we can also have the following corollary.
\begin{myCoro}
In the one-sided matching-based reverse auction of FL service market, each data-seller and UAV-seller is individual rational and truthful.
\end{myCoro}
\begin{proof}
The proof is the same as the proof in Section IV and is omitted here.
\end{proof}

Although the truthfulness and individual rationality of the proposed one-sided matching-based reverse auction have been proved, considering the selfishness of buyers and sellers, matching stability is critical to ensure the stability and efficiency of reverse auction. We first present the definition of stability of reverse auction in Definition \ref{Def-STA}.
\begin{myDef}\label{Def-STA}
(Stability of reverse auction): The proposed one-sided matching-based reverse auction is said to be stable if no buyer or sellers have incentives to deviate from the auction result.
\end{myDef}

\begin{myTheo}
The proposed one-sided matching-based reverse auction of FL service market is stable.
\end{myTheo}
\begin{proof}
First, for sellers, based on the mechanism of reverse auction, once a seller wins the auction, it can obtain non-negative revenue and this seller cannot refuse to deliver services. Besides, if any buyer $l$ is matched to the seller pair $(i,k)$, each buyer has two choices. If buyer $l$ quits, then its service requirement cannot be fulfilled and thus get zero revenue, which clearly is not a dominant strategy. When the buyer intends to replace its auction result to obtain higher revenue, however, which is not possible, since the matching result is determined by the preference list of $\mathbf{T_{A}}$ which maximizes the overall revenue. Thus, both buyers and sellers are stable in the proposed one-sided matching-based reverse auction.
\end{proof}

\begin{myRem}
The winning buyers are generally truthful and individually rational in the proposed reverse auction.
\end{myRem}
\begin{proof}
Since buyers cannot determine the final trading prices in the proposed reverse auction, the corresponding truthfulness and individual rationality of buyers are often overlooked, e.g., in \cite{TMCAuction} and \cite{NP-MASS}. This part briefly discusses these two properties of winning buyers. First, submits a service requirement bid lower than a buyer's true need, will definitely fail to meet its requirement. Besides, if a buyer reports a service requirement bid higher than its true need, the corresponding payment may increase since the payment relies heavily on the corresponding service requirements. Apparently, each buyer has no incentive to be untruthful in our proposed reverse auction model.

Individual rationality of winning buyers can be analyzed from the following two views. First, in the VCG-based reverse auction, since $F(x_{l,m,n}^{*})-F_{\setminus(i,k)}(y_{l,m,n}^{*})\geq 0$ always holds (see Theorem 2), we can conclude that if a buyer wins the auction, it can obtain non-negative revenue. More intuitively, a buyer can be selected as a winner iff it will not decrease the overall revenue. Then, in the proposed one-sided matching-based reverse auction, we build the preference list based on the preference value as defined in (\ref{e22}), and delete the buyer-seller pair associated with any negative preference value. Thus, the final winning buyers can always get non-negative revenues.
\end{proof}

\section{Simulation and Performance evaluation}
This section conducts comprehensive simulations to evaluate the feasibility of our proposed reverse auction mechanisms. Notably, this paper mainly focuses on the trading in the multiple FL services trading market based on general evaluation functions (e.g., cost function, valuation function). Besides, the proposed algorithms are executed before the specific FL training starts, which means they are irrelevant to specific FL algorithms and ML models (as also supported by existing works \cite{FedIncen-1,MultiFL-IOT,MultiFL-TWC},\cite{TMCAuction}) and can be applied in any distributed learning schemes. To this end, numerical simulations based on general assumptions of properties of participants are sufficient to verify our proposed Algorithms. Specifically, the proposed Algorithm 1 and Algorithm 2 are abbreviated as "Opt" and "Subopt" for notational simplicity. Moreover, to better evaluate performance gains achieved by the proposed algorithms, while considering the characteristics of problem given in (\ref{e10}), four heuristic methods are considered as baselines\cite{LiwangTMC}:
\begin{itemize}

 \item \textbf{Fragmental Optimization Genetic Algorithm (FOGA)}: FOGA\cite{FOGA} is a heuristic algorithm that can be used to solve tripartite matching problems, e.g., problem (\ref{e10}), which is a combination of fragmental optimization and genetic algorithm.

  \item \textbf{High Valuation Preferred Method (HVPM)}: In HVPM, a seller pair is matched to a buyer is of the highest value, based on (\ref{e5}), under constraints (10b)-(10e), until all the buyers are assigned to feasible seller pairs.

  \item \textbf{Low Cost Preferred Method (LCPM)}: Similar to HVPM, in LCPM, a seller pair is matched to a buyer is of the lowest cost, under constraints (10b)-(10e), until all the buyers are assigned to feasible seller pairs.

  \item\textbf{Random Sampling-Based Method (RSBM)}: In RSBM, a seller pair is randomly selected for each buyer, under constraints (10b)-(10e), until all the buyers are assigned to feasible seller pairs.
\end{itemize}
Notably, buyers may fail to be matched to feasible seller pairs due to factors such as the conflicts among buyers, and insufficient number of sellers.

\begin{figure}[t]
	\centering
	\includegraphics[width=3.3in]{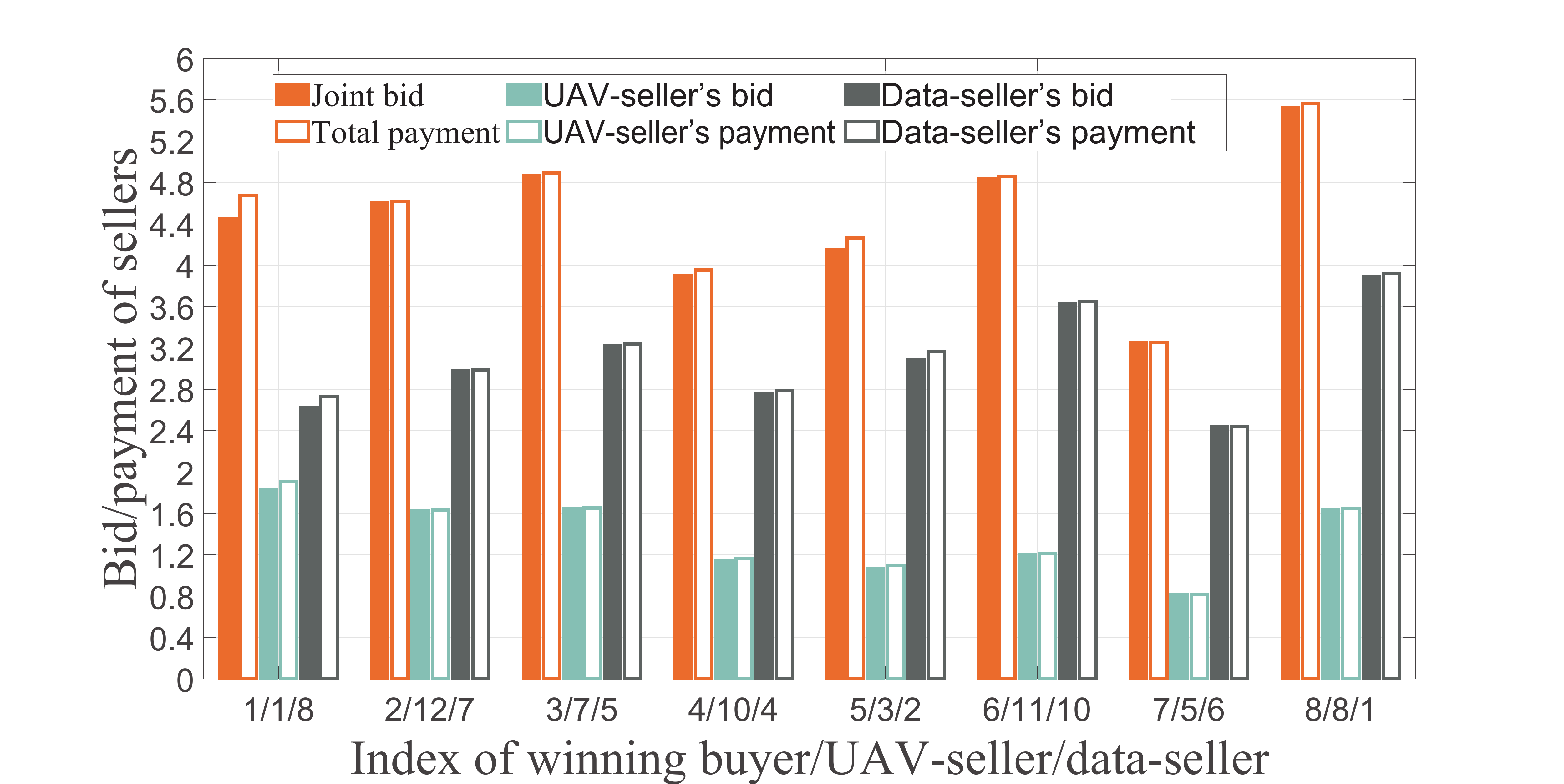}
	\caption{Individual rationality of seller-pairs and individual sellers upon considering 8 buyers, 10 UAV-sellers and 12 data-sellers.}
	\label{fig4}
\end{figure}

\begin{table*}[t]
\centering
\begin{center}\renewcommand\arraystretch{1.1}
     \caption{Detailed performance of the proposed one-sided matching-based reverse auction upon considering 8 buyers, 10 UAV-sellers and 12 data-sellers.}
     \label{tab3}
     \setlength{\tabcolsep}{0.1mm}{
\begin{tabular}{|c|c|c|c|c|c|c|c|c|c|}
\hline
Winning pair& UAV's bid&DO' bid&Joint bid&Total payment&Seller pair's revenue&UAV's payment&UAV's revenue&DO's payment&DO's revenue\\
\hline
1/1/8& 1.8304 & 2.6219&4.4523& 4.6337&0.1814&1.9066 &0.0762 &2.7271&0.1052 \\
\hline
2/12/7& 1.6292 & 2.9789&4.0681& 4.6187&0.0106&1.6329 &0.0037 &2.9858&0.0069 \\
\hline
3/7/5&  1.6438& 3.2232 & 4.8670& 4.8910& 0.0240 & 1.6519&0.0081 &3.2391&0.0159\\
\hline
4/10/4& 1.1473 & 2.7552&3.9025& 3.9537&0.0502&1.1624 & 0.0151 &2.7913&0.0361 \\
\hline
5/3/2&  1.0655 & 3.0868&4.1523&4.2638&0.1115& 1.0941 &0.0286 & 3.1697& 0.0829\\
\hline
6/11/10&  1.2048 & 3.6317&4.8365&4.8609&0.0244& 1.2109 &0.0061 & 3.6500& 0.0183\\
\hline
7/5/6& 0.8124& 2.4433&3.2557&3.2562&0.0005& 0.8125 &0.0001 & 2.4437& 0.004\\
\hline
8/8/1& 1.6318 & 3.8899&5.5217& 5.5673&0.0456&1.6453&0.0135& 3.9220&0.0321\\
\hline
\end{tabular}}
\end{center}
\end{table*}

\begin{figure*}[t]
    \centering
    \subfigure[] {\includegraphics[width=2.3in,angle=0]{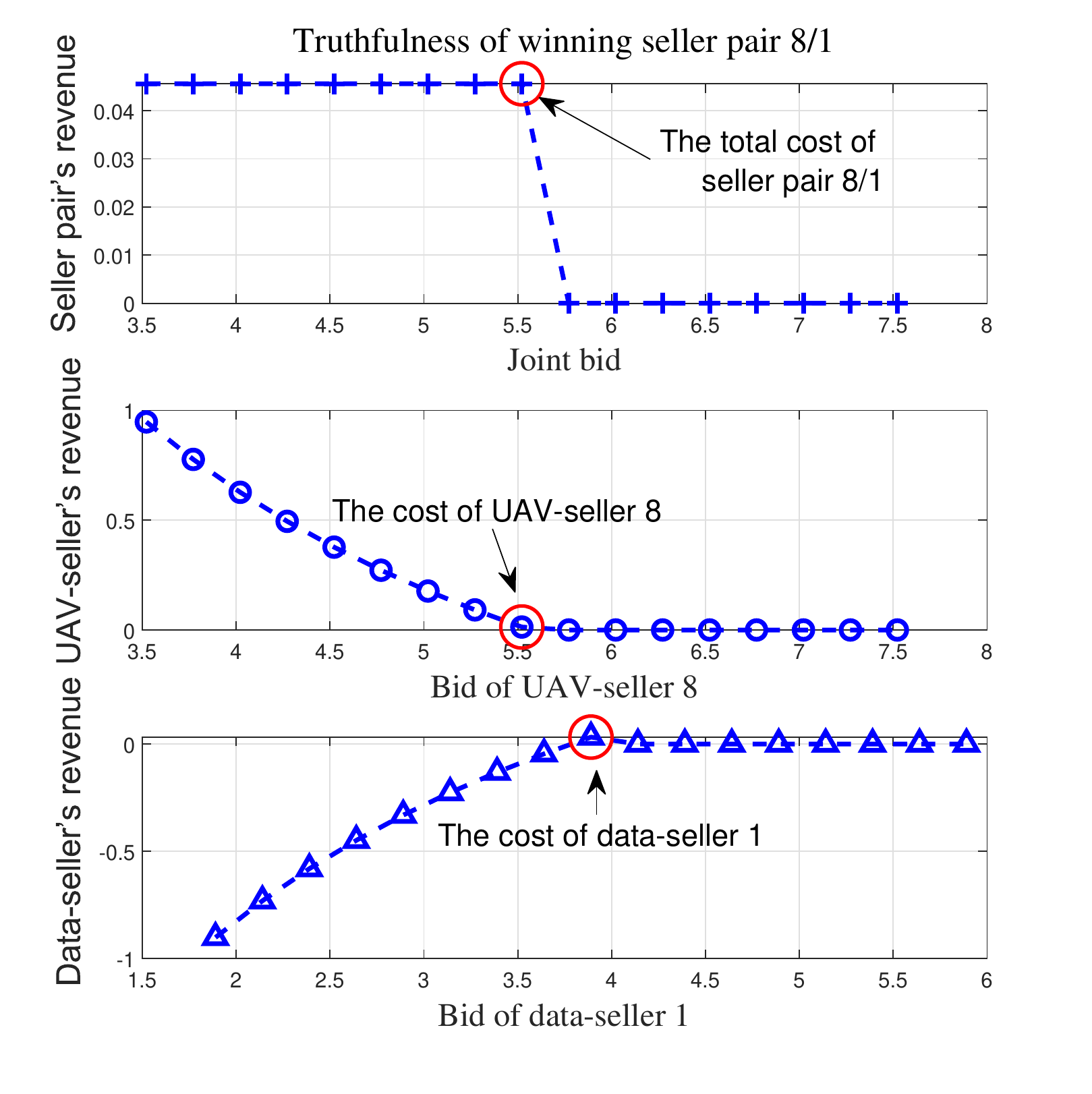}}
    \subfigure[] {\includegraphics[width=2.3in,angle=0]{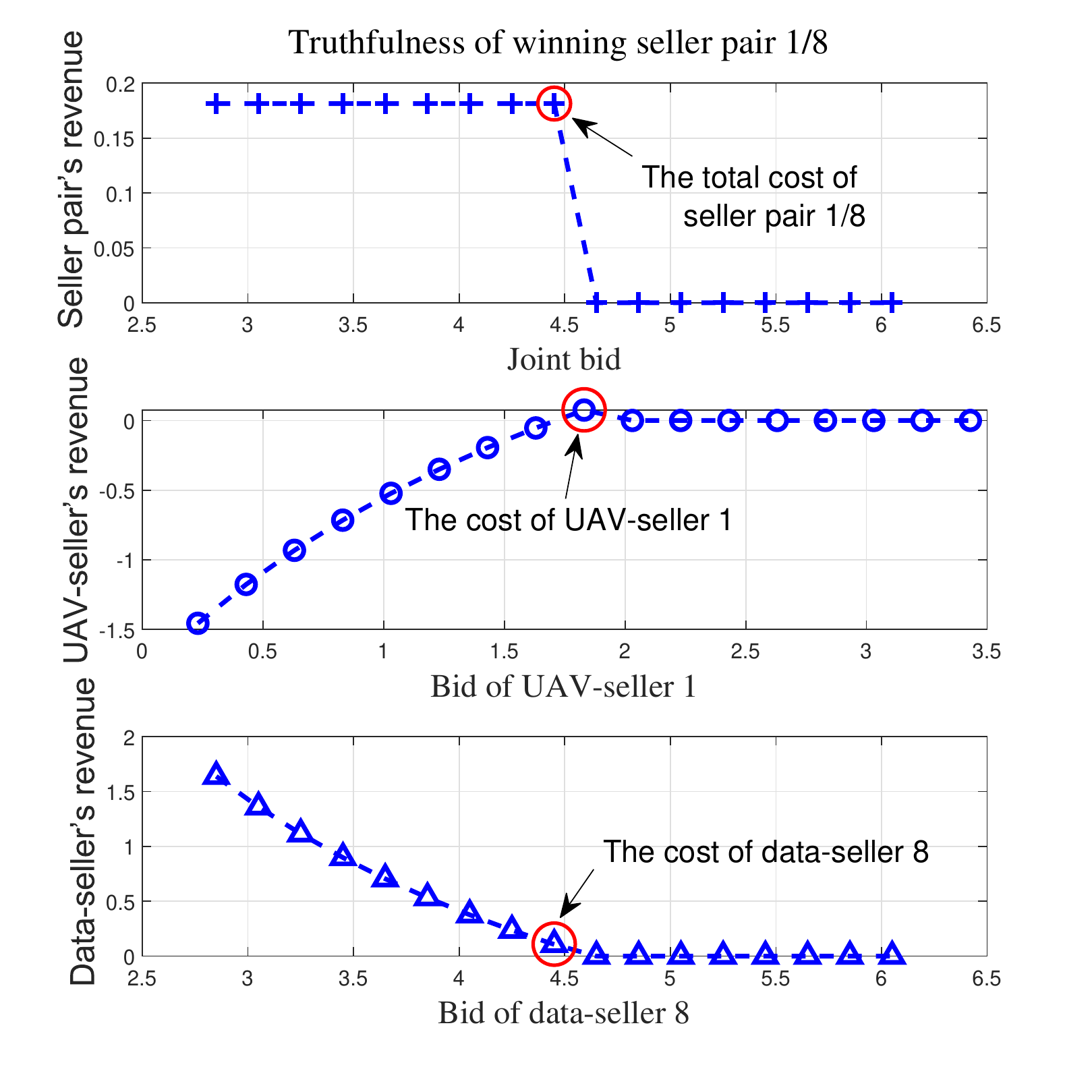}}
    \subfigure[] {\includegraphics[width=2.3in,angle=0]{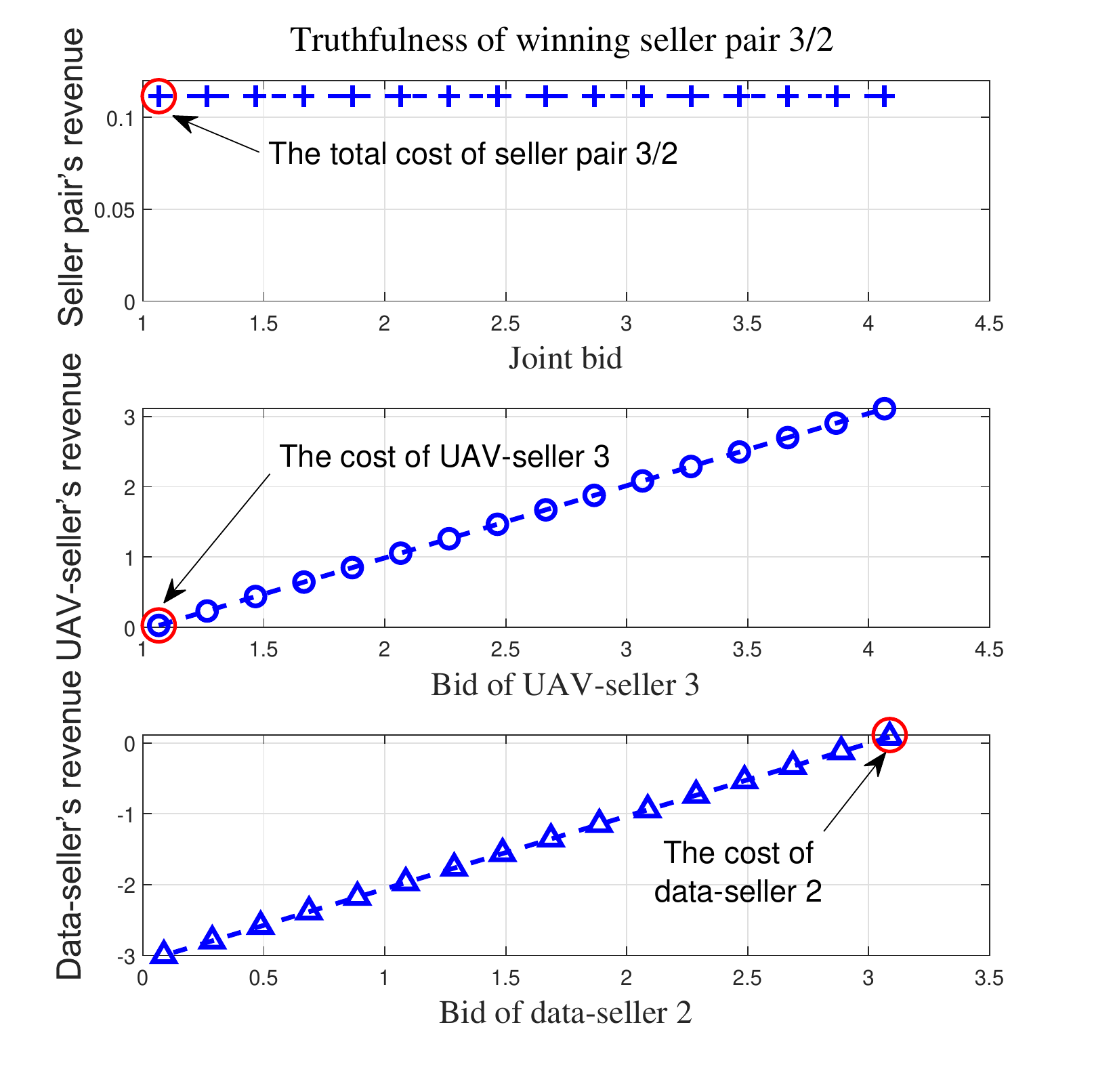}}
    \caption{Truthfulness of seller-pairs and individual sellers: a) UAV-seller bids truthfully and data-seller bids untruthfully; b) UAV-seller bids untruthfully and data-seller bids truthfully; c) both sellers bid untruthfully when the joint bid remains truthful.}
    \label{fig5}
\end{figure*}

\subsection{Simulation settings}
For data-sellers, we assume that the normalized data size (normalized by 500 units\footnote{One unit could be 1 KB, 1 MB or 1 GB data.}) $d_{m,l}$ of each data-seller $m$ follows a uniform distribution $[10, 30]$, and the unit data cost $\sigma_{m,l}$ can be randomly selected from a uniform distribution $[0.0002, 0.0004]$. For UAV-sellers, we assume the fly distance $t_{n}$ follows the distribution $[10,100]$ meters, while the unit flying cost $\lambda_{n}$ follows the distribution $[0.02, 0.05]$. For simplicity, we mainly consider the UAV communication delay during the model transmission as the estimation of UAV-service cost function \cite{TMCAuction}\footnote{More complicated service costs can be replaced by considering specific scenario or communication protocols as future work.}. Specifically, we assume the model size of each buyer follows a uniform distribution $[100, 500]$ KB, while the communication rate between UAV-seller $n$ and data-seller $m$ follows a uniform distribution $[100, 300]$ KB/s\cite{MultiFL-IOT}, so that the model transmission delay (namely, service cost of UAV) can be calculated.  For buyers, the valuation function of each buyer is supposed to be calculated by a log function $\alpha_{1}*log(1+\alpha_{2}*d)$ according to \cite{FedIncen-1}, where $\alpha_2=1$ and $\alpha_1$ follows a uniform distribution $[8, 12]$, which can be different from various buyers. We set $d_{l}=5000$ for all the buyers. We conduct the simulation with MATLAB 2021b on Intel(R) Core(M) i7-11700F@2.5 GHz, and the simulation results demonstrated in the following sections are the results averaged over 1000 simulations (auctions), unless otherwise stated.

\subsection{Performance of running time}
The running time performance comparison is detailed by Table \ref{tab2}, upon considering different number of buyers, UAV-sellers, and data-sellers. For example, let 2/4/6 denote the problem of 2 buyers, 4 UAV-sellers and 6 data-sellers. Notably, the running time of the proposed Opt algorithm is about $10^2$ to $10^6$ times that of the other five methods. Since a large number of enumeration and permutation calculations are required during the execution process of Opt algorithm, which thus causes huge memory pressure and makes it computationally intractable and unpractical in implementation. Moreover, it can be observed that the proposed Subopt algorithm has the same or lower level regarding running time in comparison with baseline methods, which proves the corresponding computational efficiency, and thus can achieve acceptable running time for various problem sizes.

\subsection{Performance of total revenue of buyers}
The comparisons on the total revenue of buyers, i.e., the value of $F(x_{l,m,n})$ defined in (\ref{e9}), among six methods upon considering various problem sizes, are demonstrated in Fig. 3. Firstly, Fig. 3 indicates that the proposed Subopt algorithm achieves similar or approaches to the total revenue of FOGA and the Opt algorithm, and greatly outperforms the other three baseline methods. For example, compared with Opt, the obtained total revenue is only decreased by 2.25\% via applying the proposed Subopt algorithm and decreased by 1.5\%, 25.64\%, 49.70\%, and 47.30\% when considering FOGA, HVPM, LCPM, and RSBM, respectively, under problem size 6/8/10. Although FOGA achieves slightly larger revenue than the Subopt algorithm under the cost of longer running time, it fails to guarantee the properties of truthfulness, individual rationality, and computational efficiency. Revisit the above-mentioned running time performance, the proposed Subopt algorithm can achieve a satisfying trade-off between running time and total revenue, in comparison with FOGA. Then, Fig. 3(b), Fig. 3(c), and Fig. 3(d) illustrate the performance of the total revenue versus the number of buyers, UAV-sellers, and data-sellers, respectively, when keeping the number of other two (of the three) parties fixed at 5. Obviously, the total revenue of different methods increases with the increasing number of participants, at different growth rates. Interestingly, the growth rates of Opt, Subopt and FOGA slowed down significantly after 5 (the value of x-axis), while other methods can still maintain a slow growth. This is because as the number of participants increases, other methods can obtain better suboptimal solutions thanks to a larger searching space.

\subsection{Performance of economical properties}
Since the properties of the VCG-based reverse auction have been extensively verified in existing works, we thus focus on the properties (truthfulness and individual rationality) of the proposed Subopt algorithm. Specifically, we conduct the simulations under 8 buyers, 10 UAV-sellers, and 12 data-sellers. Performance on individual rationality is shown in Fig. \ref{fig4}, with the joint bid and the total payment of winning seller pairs (8 in this figure), as well as individual bids and the corresponding payment of each UAV/data seller. For example, 2/12/7 means buyer 2 is matched to the winning UAV-seller 12 and data-seller 7. Notably, each winning seller can obtain a final payment no less than its bid, which proves the individual rationality of both seller pairs and individual sellers as given in \textbf{Theorem 5} and \textbf{Corollary 3}. To achieve better analysis, detailed performance of each winning seller pair is shown by Table \ref{tab3}.

Fig. \ref{fig5} illustrates the truthfulness of seller pairs and individual sellers. As discussed in subsection C of Section IV, three seller pairs (seller pair 8/1, 1/8, and 3/2) in Table \ref{tab3} are taken as examples. In Fig. 5(a), UAV-seller 8 bids truthfully at 1.6318 while data-seller 1 bids untruthfully from 1.8899 to 5.8899. Notably, if data-seller 1 bids smaller than its cost, the revenue of seller pair remains the same at 0.0456 and becomes zero when data-seller 1 overbids, while data-seller 1 only obtains positive revenue 0.0321 when the bid equals its cost 3.8899 (marked by a red circle), which indicates that any individual seller or seller pair cannot obtain higher revenue by bidding untruthfully, similar conclusion can also be concluded from Fig. 3(b). In Fig. 3(c), UAV-seller 3 overbids from 1.0655 to 4.0655, and data-seller 2 underbids from 3.0688 to 0.0868, and the joint bid remains truthful. It can be observed that the seller pair's revenue remains unchanged, and the UAV-seller's revenue increases with the increase of its bid, while data-seller 2 only obtains negative revenue by underbidding, which is consistent with the analysis in \textbf{Corollary 2}. Finally, we can conclude from Fig. \ref{fig5} that the proposed Subopt algorithm greatly holds the property of truthfulness.

\section{Conclusion}
In this paper, we study a novel multiple FL services trading problem among buyers, data-sellers and UAV-sellers, in a UAV-aided network based on a well-designed reverse auction. A 0-1 integer programming problem is formulated to maximize the overall revenue of buyers. An interesting concept of seller pair and joint bid is proposed to facilitate the trading among these three parties. We first propose a VCG-based reverse auction mechanism to obtain the optimal solutions which, however, is computationally intractable. We then propose a computation-efficient one-sided matching-based reverse auction mechanism to obtain suboptimal solutions that approach to optimal ones, upon considering a large number of participants. Significant properties such as truthfulness and individual rationality are comprehensively analyzed for both mechanisms. Finally, extensive simulation results demonstrate the effectiveness of our proposed algorithms as compared with four baseline methods.




\section*{Acknowledgment}

This work is supported in part by the National Natural Science Foundation of China (grant nos. 61971365, 61871339, 62171392), Digital Fujian Province Key Laboratory of IoT Communication, Architecture and Safety Technology (grant no. 2010499), the State Key Program of the National Natural Science Foundation of China (grant no. 61731012), the Natural Science Foundation of Fujian Province of China No. 2021J01004.

\ifCLASSOPTIONcaptionsoff
  \newpage
\fi


\begin{thebibliography}{1}
\addtolength{\itemsep}{-0.1em}
\bibitem{6GNetwk-1}
W. Saad, M. Bennis and M. Chen, ``A Vision of 6G Wireless Systems: Applications, Trends, Technologies, and Open Research Problems,'' \textit{IEEE Netw.}, vol. 34, no. 3, pp. 134-142, Jun. 2020.
\bibitem{new-add-zhang1}
S. Zhang, H. Zhang, and L. Song,``Beyond D2D: Full Dimension UAV-to-Everything Communications in 6G,'' \textit{IEEE Trans. Veh. Technol.}, vol. 69, no. 6, pp. 6592-6602, Jun. 2020.
\bibitem{6GNetwk-2}
W. -C. Chien et al., ``Intelligent Architecture for Mobile HetNet in B5G,'' \textit{IEEE Netw.}, vol. 33, no. 3, pp. 34-41, Jun. 2019.
\bibitem{AliMag}
S. Hosseinalipour, C. G. Brinton, V. Aggarwal, H. Dai and M. Chiang, ``From Federated to Fog Learning: Distributed Machine Learning over Heterogeneous Wireless Networks," \textit{IEEE Commun. Mag.}, vol. 58, no. 12, pp. 41-47, Dec. 2020.
\bibitem{GoogleFL2016}
J. Konecny, H. B. McMahan, F. X. Yu, P. Richtarik, A. T. Suresh, and D. Bacon, ``Federated learning: Strategies for improving communication efficiency,” arXiv preprint arXiv:1610.05492, 2016.
\bibitem{FedCS-1}
Y. Deng et al., ``AUCTION: Automated and Quality-Aware Client Selection Framework for Efficient Federated Learning," \textit{IEEE Trans. Parallel Distrib. Syst.}, vol. 33, no. 8, pp. 1996-2009, Aug. 2022.
\bibitem{GsongNet21}
Y. Zhan, P. Li, S. Guo and Z. Qu, ``Incentive Mechanism Design for Federated Learning: Challenges and Opportunities," \textit{IEEE Netw.}, vol. 35, no. 4, pp. 310-317, July/August 2021.
\bibitem{LaaS}
M. Ribeiro, K. Grolinger and M. A. M. Capretz, ``MLaaS: Machine Learning as a Service," in \textit{Proc. 2015 IEEE 14th International Conference on Machine Learning and Applications (ICMLA)}, Miami, FL, USA, Dec. 2015.
\bibitem{FedAvg}
 H. B. McMahan et al., ``Communication-Efficient Learning of Deep Networks from Decentralized Data,'' in \textit{Proc. Int’l. Conf. Artificial Intelligence and Statistics}, Apr. 2017
\bibitem{FedCS-2}
T. Huang, W. Lin, W. Wu, L. He, K. Li and A. Y. Zomaya, ``An Efficiency-Boosting Client Selection Scheme for Federated Learning With Fairness Guarantee," \textit{IEEE Trans. Parallel Distrib. Syst.}, vol. 32, no. 7, pp. 1552-1564, Jul. 2021.
\bibitem{FedRM-1}
Q. Zeng, Y. Du, K. Huang and K. K. Leung, ``Energy-Efficient Resource Management for Federated Edge Learning with CPU-GPU Heterogeneous Computing," \textit{IEEE Trans. Wireless Commun.}, vol. 20, no. 12, pp. 7947-7962, Dec. 2021.
\bibitem{FedRM-2}
V. -D. Nguyen, S. K. Sharma, T. X. Vu, S. Chatzinotas and B. Ottersten, ``Efficient Federated Learning Algorithm for Resource Allocation in Wireless IoT Networks," \textit{IEEE Internet Things J.}, vol. 8, no. 5, pp. 3394-3409, Mar. 2021.
\bibitem{FedHier-1}
L. Liu, J. Zhang, S. H. Song and K. B. Letaief, ``Client-Edge-Cloud Hierarchical Federated Learning," in \textit{Proc. 2020 IEEE International Conference on Communications (ICC)}, Dublin, Ireland, Jun. 2020.
\bibitem{FedHier-2}
A. Hashemi, A. Acharya, R. Das, H. Vikalo, S. Sanghavi and I. S. Dhillon, ``On the Benefits of Multiple Gossip Steps in Communication-Constrained Decentralized Federated Learning," \textit{IEEE Trans. Parallel Distrib. Syst.}, in press, doi: 10.1109/TPDS.2021.3138977.
\bibitem{FedIncen-1}
Y. Zhan, P. Li, Z. Qu, D. Zeng and S. Guo, ``A Learning-Based Incentive Mechanism for Federated Learning," \textit{IEEE Internet Things J.}, vol. 7, no. 7, pp. 6360-6368, Jul. 2020.
\bibitem{MultiFL-IOT}
D. Chen et al., ``Matching-Theory-Based Low-Latency Scheme for Multitask Federated Learning in MEC Networks," \textit{IEEE Internet Things J.}, vol. 8, no. 14, pp. 11415-11426, Jul. 2021.
\bibitem{MultiFL-TWC}
J. Xu, H. Wang and L. Chen, ``Bandwidth Allocation for Multiple Federated Learning Services in Wireless Edge Networks,"  \textit{IEEE Trans. Wireless Commun.}, vol. 21, no. 4, pp. 2534-2546, Apr. 2022.
\bibitem{MultiFL-TMC}
M. N. H. Nguyen, N. H. Tran, Y. K. Tun, Z. Han and C. S. Hong, ``Toward Multiple Federated Learning Services Resource Sharing in Mobile Edge Networks,"  \textit{IEEE Trans. Mobile Comput.}, in press, doi: 10.1109/TMC.2021.3085979.
\bibitem{DataTrade-TIFS}
W. Dai, C. Dai, K. -K. R. Choo, C. Cui, D. Zou and H. Jin, ``SDTE: A Secure Blockchain-Based Data Trading Ecosystem," \textit{IEEE Trans. Inf. Forensics Security.}, vol. 15, pp. 725-737, Jul. 2020.
\bibitem{NewaddTPDS-1}
Z. Zhou, F. Liu, S. Chen and Z. Li, ``A Truthful and Efficient Incentive Mechanism for Demand Response in Green Datacenters," \textit{IEEE Trans. Parallel Distrib. Syst.}, vol. 31, no. 1, pp. 1-15, Jan. 2020.
\bibitem{NewaddTPDS-2}
Y. Jiao, P. Wang, D. Niyato and K. Suankaewmanee, ``Auction Mechanisms in Cloud/Fog Computing Resource Allocation for Public Blockchain Networks," \textit{IEEE Trans. Parallel Distrib. Syst.}, vol. 30, no. 9, pp. 1975-1989, Sept. 2019.
\bibitem{NewaddTPDS-3}
T. Bahreini, H. Badri and D. Grosu, ``Mechanisms for Resource Allocation and Pricing in Mobile Edge Computing Systems," \textit{IEEE Trans. Parallel Distrib. Syst.}, vol. 33, no. 3, pp. 667-682, Mar. 2022.

\bibitem{New-review3}
X. Gao, P. Wang, D. Niyato, K. Yang and J. An, ``Auction-Based Time Scheduling for Backscatter-Aided RF-Powered Cognitive Radio Networks," \textit{IEEE Trans. Wireless Commun.}, vol. 18, no. 3, pp. 1684-1697, Mar. 2019.

\bibitem{Auction-FL-TWC}
T. H. Thi Le et al., ``An Incentive Mechanism for Federated Learning in Wireless Cellular Networks: An Auction Approach," \textit{IEEE Trans. Wireless Commun.}, vol. 20, no. 8, pp. 4874-4887, Aug. 2021.
\bibitem{new-add-zhang2}
H. Zhang, L. Song, and Z. Han, ``Unmanned Aerial Vehicle Applications over Cellular Networks for 5G and Beyond,'' New York, NY, USA: Springer, 2020.

\bibitem{FL-UAV-Network}
W. Y. B. Lim et al., ``UAV-assisted Communication Efficient Federated Learning in the Era of the Artificial Intelligence of Things," \textit{IEEE Netw.}, vol. 35, no. 5, pp. 188-195, Sept./Oct. 2021.
\bibitem{21-TITS-MC}
W. Y. B. Lim et al., ``Towards Federated Learning in UAV-Enabled Internet of Vehicles: A Multi-Dimensional Contract-Matching Approach," \textit{IEEE Trans. Intell. Transport. Syst.}, vol. 22, no. 8, pp. 5140-5154, Aug. 2021.
\bibitem{TITS-A-C}
J. S. Ng et al., ``Joint Auction-Coalition Formation Framework for Communication-Efficient Federated Learning in UAV-Enabled Internet of Vehicles," \textit{IEEE Trans. Intell. Transport. Syst.}, vol. 22, no. 4, pp. 2326-2344, Apr. 2021.
\bibitem{FLSurvey-1}
W. Y. B. Lim et al., ``Federated Learning in Mobile Edge Networks: A Comprehensive Survey," \textit{IEEE Commun. Surveys Tuts.}, vol. 22, no. 3, pp. 2031-2063, thirdquarter 2020.
\bibitem{FLSurvey-2}
S. Abdulrahman, H. Tout, H. Ould-Slimane, A. Mourad, C. Talhi and M. Guizani, ``A Survey on Federated Learning: The Journey From Centralized to Distributed On-Site Learning and Beyond," \textit{IEEE Internet Things J.}, vol. 8, no. 7, pp. 5476-5497, Apr. 2021.
\bibitem{FLSurvey-3}
D. C. Nguyen, M. Ding, P. N. Pathirana, A. Seneviratne, J. Li and H. Vincent Poor, ``Federated Learning for Internet of Things: A Comprehensive Survey," \textit{IEEE Commun. Surveys Tuts.}, vol. 23, no. 3, pp. 1622-1658, thirdquarter 2021.
\bibitem{TMCAuction}
Y. Jiao, P. Wang, D. Niyato, B. Lin and D. I. Kim, ``Toward an Automated Auction Framework for Wireless Federated Learning Services Market," \textit{IEEE Trans. Mobile Comput.}, vol. 20, no. 10, pp. 3034-3048, Oct. 2021.
\bibitem{FL-UAV-TVT}
H. Zhang and L. Hanzo, ``Federated Learning Assisted Multi-UAV Networks," \textit{IEEE Trans. Veh. Technol.}, vol. 69, no. 11, pp. 14104-14109, Nov. 2020.

\bibitem{TMC20-BudgetConstraint}
C. Dai, X. Wang, K. Liu, D. Qi, W. Lin and P. Zhou, ``Stable Task Assignment for Mobile Crowdsensing with Budget Constraint," \textit{IEEE Trans. Mobile Comput.}, vol. 20, no. 12, pp. 3439-3452, Dec. 2021.
\bibitem{LiwangIoT2019}
M. Liwang, S. Dai, Z. Gao, Y. Tang and H. Dai, ``A Truthful Reverse-Auction Mechanism for Computation Offloading in Cloud-Enabled Vehicular Network," \textit{IEEE Internet Things J.}, vol. 6, no. 3, pp. 4214-4227, Jun. 2019.
\bibitem{U-cost-1}
Y. Luo, X. Huang, J. Yang, F. Wu and S. Leng, ``Auction Mechanism-based Multi-type Task Planning for Heterogeneous UAVs Swarm," in \textit{Proc. 2020 IEEE 20th International Conference on Communication Technology (ICCT)}, Oct. 2020.
\bibitem{U-cost-2}
H. Khdr, M. Shafique, S. Pagani, A. Herkersdorf and J. Henkel, ``Combinatorial Auctions for Temperature-Constrained Resource Management in Manycores," \textit{IEEE Trans. Parallel Distrib. Syst.}, vol. 31, no. 7, pp. 1605-1620, Jul. 2020.
\bibitem{LiwangTWC22}
M. Liwang, R. Chen, X. Wang, and X. Shen, ``Unifying Futures and Spot Market: Overbooking-Enabled Resource Trading in Mobile Edge Networks," \textit{IEEE Trans. Wireless Commun.}, in press, doi: 10.1109/TWC.2022.3141094.

\bibitem{INFOCOM03}
S. Zhong, J. Chen and Y. R. Yang, "Sprite: A Simple, Cheat-Proof, Credit-Based System for Mobile Ad-Hoc Networks," in \textit{Proc. 22nd Annu. Joint Conf. IEEE Comput. Commun. Soc.}, San Francisco, CA, USA, Apr. 2003.

\bibitem{STAR}
Z. Zheng, Y. Gui, F. Wu and G. Chen, ``STAR: Strategy-Proof Double Auctions for Multi-Cloud, Multi-Tenant Bandwidth Reservation," \textit{IEEE Trans. Comput.}, vol. 64, no. 7, pp. 2071-2083, Jul. 2015.
\bibitem{NP-TNSM}
W. Borjigin, K. Ota and M. Dong, ``In Broker We Trust: A Double-Auction Approach for Resource Allocation in NFV Markets," \textit{IEEE Trans. Netw. Service Manag.}, vol. 15, no. 4, pp. 1322-1333, Dec. 2018.
\bibitem{NP-MASS}
A. R. Khamesi and S. Silvestri, ``Reverse Auction-based Demand Response Program: A Truthful Mutually Beneficial Mechanism," in \textit{Proc. 2020 IEEE 17th International Conference on Mobile Ad Hoc and Sensor Systems (MASS)}, Delhi, India, Dec. 2020.
\bibitem{AuctionMatching}
Patel, Y.S., Malwi, Z., Nighojkar, A. et al. ``Truthful Online Double Auction Based Dynamic Resource Provisioning for Multi-Objective Trade-Offs in IaaS Clouds". \textit{Cluster Comput.},vol.24, pp.1855–1879, Jan. 2021.
\bibitem{HyperMatch}
Q. Wei, W. Sun, B. Bai, L. Wang, E. G. Ström and M. Song, ``Resource Allocation for V2X Communications: A Local Search Based 3D Matching Approach," in \textit{Proc. 2017 IEEE International Conference on Communications (ICC)}, May 2017.
\bibitem{HierDA}
W. Tang and R. Jain, ``Hierarchical Auction Mechanisms for Network Resource Allocation," \textit{IEEE J. Select. Areas Commun.}, vol. 30, no. 11, pp. 2117-2125, Dec. 2012.
\bibitem{TierDA}
M. Stübs, W. Posdorfer and S. Momeni, ``Blockchain-Based Multi-Tier Double Auctions for Smart Energy Distribution Grids," in \textit{Proc. 2020 IEEE International Conference on Communications Workshops (ICC Workshops)}, May 2020.
\bibitem{McAfee}
R. McAfee, ``A Dominant Strategy Double Auction", \textit{J. Econ. Theory}, vol. 56, no. 2, pp. 434-450, 1992.
\bibitem{MIDA-18}
Segal-Halevi, E., Hassidim, A., and Aumann, Y. ``Double Auctions in Markets for Multiple Kinds of Goods", in \textit{Proc. 2018 IJCAI}, Stockholm, Sweden. Jul. 2018.
\bibitem{MUDA-18}
Segal-Halevi, E., Hassidim, A., and Aumann, Y. `` MUDA: A Truthful Multi-Uint Double Auction Mechanism", in \textit{Proc. Thirty-Second AAAI Conference on Artificial Intelligence (AAAI)}, New Orleans, USA. Feb. 2018.
\bibitem{LiwangTMC}
Z. Gao, M. Liwang, S. Hosseinalipour, H. Dai and X. Wang, ``A Truthful Auction for Graph Job Allocation in Vehicular Cloud-assisted Networks," \textit{IEEE Trans. Mobile Comput.}, in press, doi: 10.1109/TMC.2021.3059803.
\bibitem{FOGA}
G. Huang and Andrew L., ``A Hybrid Genetic Algorithm for Three-Index Assignment Problem,'' in \textit{Proc. CEC '03.}, Canberra, ACT, Australia, Dec. 2003.




\end{thebibliography}
\end{document}